\documentclass[titlepage, twocolumn, aps, prd, longbibliography, nofootinbib, superscriptaddress]{revtex4}

\usepackage{appendix}
\usepackage{graphicx}
\usepackage{amsmath}
\usepackage{amsfonts}
\usepackage{dsfont}
\usepackage{mathtools}
\usepackage{xcolor}
\usepackage[caption=false]{subfig}

\usepackage[pdftex,colorlinks=true]{hyperref}
\hypersetup{linkcolor=blue,citecolor = blue,urlcolor=blue}

\newcommand{\RNum}[1]{\uppercase\expandafter{\romannumeral #1\relax}}

\newcommand\numberthis{\addtocounter{equation}{1}\tag{\theequation}}

\newcommand{\ket}[1]{| #1 \rangle}
\newcommand{\bra}[1]{\langle #1 |}

\DeclarePairedDelimiter\abs{\lvert}{\rvert}%
\DeclarePairedDelimiter\norm{\lVert}{\rVert}%

\makeatletter
\let\oldabs\abs
\def\abs{\@ifstar{\oldabs}{\oldabs*}}
\let\oldnorm\norm
\def\norm{\@ifstar{\oldnorm}{\oldnorm*}}
\makeatother

\begin{document}

\title{Control design for inhomogeneous broadening compensation in single-photon transducers}

\author{Sattwik Deb Mishra$^{*}$}
\affiliation{Ginzton Laboratory, Stanford University, 348 Via Pueblo Mall, Stanford, California 94305, USA} 
\author{Rahul Trivedi$^{*}$}
\affiliation{Max-Planck-Institute of Quantum Optics, Hans-Kopfermann-Str. 1, Garching 85748, Germany}
\author{Amir H. Safavi-Naeini}
\affiliation{Ginzton Laboratory, Stanford University, 348 Via Pueblo Mall, Stanford, California 94305, USA} 
\author{Jelena Vu\u{c}kovi\'{c}}
\affiliation{Ginzton Laboratory, Stanford University, 348 Via Pueblo Mall, Stanford, California 94305, USA}

\begin{abstract}

    A transducer of single photons between microwave and optical frequencies can be used to realize quantum communication over optical fiber links between distant superconducting quantum computers. A promising scalable approach to constructing such a transducer is to use ensembles of quantum emitters interacting simultaneously with electromagnetic fields at optical and microwave frequencies. However, inhomogeneous broadening in the transition frequencies of the emitters can be detrimental to this collective action. In this article, we utilise a gradient-based optimization strategy to design the temporal shape of the laser field driving the transduction system to mitigate the effects of inhomogeneous broadening. We study the improvement of transduction efficiencies as a function of inhomogeneous broadening in different single-emitter cooperativity regimes and correlate it with a restoration of superradiance effects in the emitter ensembles. Furthermore, to assess the optimality of our pulse designs, we provide certifiable bounds on the design problem and compare them to the achieved performance.

\end{abstract}


\maketitle
\def\thefootnote{}\footnotetext{$^*$ These authors contributed equally to this work.}

\section*{Introduction} \label{sec:intro}

Current superconducting quantum systems are able to achieve non-trivial quantum computational tasks \cite{aruteQuantumSupremacyUsing2019} and connecting them as nodes of a quantum internet can realize scalable, distributed quantum computing \cite{kimbleQuantumInternet2008a}. Since superconducting quantum systems operate at microwave frequencies, there are technological restrictions to directly connecting distant systems. Commercial microwave cables are dominated by thermal noise at room temperature and hence cause huge loss over long distances. On the other hand, cryo-cooled superconducting transmission lines are low loss but limited to short distances \cite{magnardMicrowaveQuantumLink2020}. Optical photons are better `flying' qubits; they can be transmitted with low loss over long distances through optical fibers. To connect superconducting quantum systems, there is a necessity to realize coherent transduction systems that can convert photons coherently and bi-directionally between microwave and optical frequencies.

Many approaches have been proposed to construct such transducers \cite{lambertCoherentConversionMicrowave2020,laukPerspectivesQuantumTransduction2020}. Microwave-to-optical transducers couple fields oscillating at the respective frequencies through a non-linear medium that can be driven externally to bridge the gap between these frequency regimes. The different types of non-linear media that have been studied so far are, electro-optic materials \cite{mckennaCryogenicMicrowavetoopticalConversion2020, holzgrafeCavityElectroopticsThinfilm2020, soltaniEfficientQuantumMicrowavetooptical2017a, tsangCavityQuantumElectrooptics2010a, ruedaElectroopticEntanglementSource2019, ruedaEfficientMicrowaveOptical2016}, magnon modes \cite{hisatomiBidirectionalConversionMicrowave2016a, evertsMicrowaveOpticalPhoton2019a, evertsUltrastrongCouplingMicrowave2020}, optomechanical systems \cite{zhongProposalHeraldedGeneration2020, wuMicrowavetoOpticalTransductionUsing2020, lauGroundStateCooling2020, jiangEfficientBidirectionalPiezooptomechanical2020, forschMicrowavetoopticsConversionUsing2020, arnoldConvertingMicrowaveTelecom2020, bagciOpticalDetectionRadio2014a, andrewsBidirectionalEfficientConversion2014a, wangUsingInterferenceHigh2012a, tianAdiabaticStateConversion2012a, hillCoherentOpticalWavelength2012a, barzanjehReversibleOpticaltoMicrowaveQuantum2012a, safavi-naeiniProposalOptomechanicalTraveling2011a, tianOpticalWavelengthConversion2010a, stannigelOptomechanicalTransducersLongDistance2010a}, and broadly, ensembles of atomic systems \cite{bartholomewOnchipCoherentMicrowavetooptical2020a, barnettTheoryMicrowaveOpticalConversion2020, welinskiElectronSpinCoherence2019, vogtEfficientMicrowavetoopticalConversion2019a, petrosyanMicrowaveOpticalConversion2019, fernandez-gonzalvoCavityenhancedRamanHeterodyne2019a, coveyMicrowavetoopticalConversionFourwave2019, hanCoherentMicrowavetoOpticalConversion2018a, gardMicrowavetoopticalFrequencyConversion2017a, kiffnerTwowayInterconversionMillimeterwave2016a, fernandez-gonzalvoCoherentFrequencyUpconversion2015, williamsonMagnetoOpticModulatorUnit2014a, obrienInterfacingSuperconductingQubits2014, hafeziAtomicInterfaceMicrowave2012, verduStrongMagneticCoupling2009a, imamogluCavityQEDBased2009a}. 

\begin{figure*}[t]
    \centering
    \includegraphics[width=\textwidth]{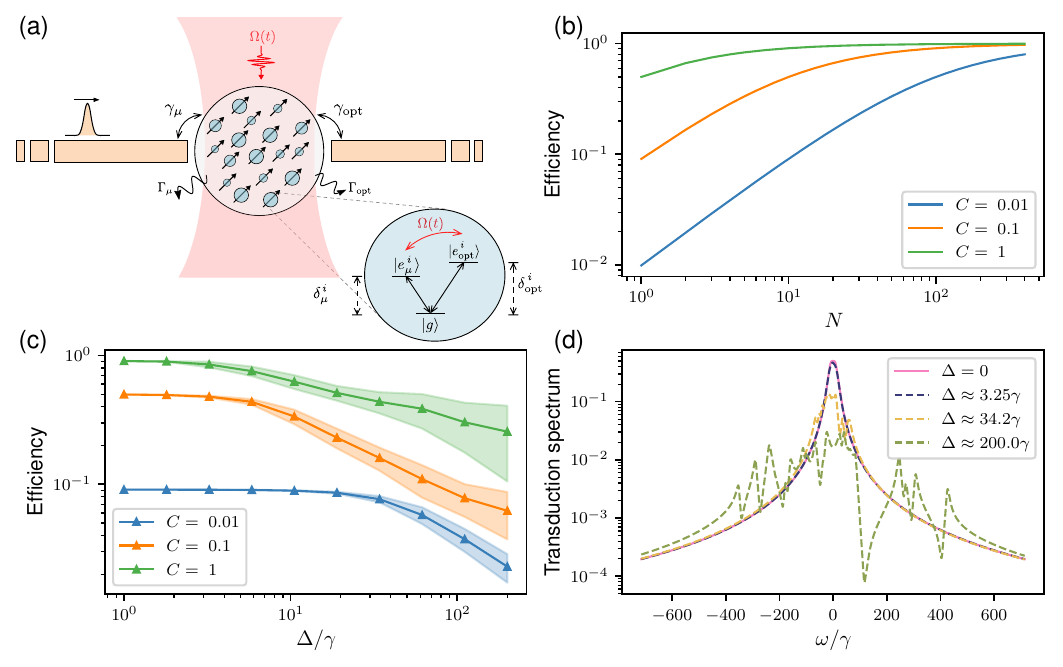}
    \caption{(a) Schematic of a three-level system ensemble-based transducer device. (b) Scaling of transduction efficiency with increasing number $(N)$ of three-level systems in a homogeneous ensemble for different cooperativities $C$ (we keep $\gamma$ fixed and vary $\Gamma$ to vary cooperativity). (c) Decrease in the transduction efficiency through randomly inhomogeneously broadened ensembles of $N = 10$ emitters with increasing inhomogeneous broadening $\Delta$ for different cooperativities $C$. For each value of the inhomogeneous broadening $\Delta$, $100$ randomly broadened ensembles are created by sampling the emitter detunings $\delta_{\mu}^{\left( i \right) }, \delta_{\text{opt}}^{\left( i \right)}$ from a Gaussian distribution with standard deviation equal to $\Delta$.  Each plot point corresponds to the mean over the 100 ensembles with inhomogeneous broadening equal to the corresponding value of $\Delta$ and the shaded regions represent the standard deviation. (d) Transduction spectra of ensembles ($N = 10, \ C = 0.1$) with varying inhomogeneous broadening $\Delta$.}
    \label{fig:1}
\end{figure*} 

Solid-state emitters (like color centers in diamond and silicon carbide and rare-earth ions doped in crystals) can have transitions coupling to both microwave and optical fields. They provide an attractive platform for implementing transducers owing to the possibility of integration with superconducting quantum systems \cite{zhuCoherentCouplingSuperconducting2011, doldHighCooperativityCouplingRareEarth2019} and scalability afforded by rapidly developing nano-fabrication techniques \cite{doryInversedesignedDiamondPhotonics2019, lukin4HsiliconcarbideoninsulatorIntegratedQuantum2020, wanLargescaleIntegrationArtificial2020}. However, single defects are often only weakly coupled to the microwave and optical fields, leading to low transduction efficiencies. An approach to overcoming this limitation is to use ensembles of such emitters coupling to the same microwave and optical channels — the coupling strength is then enhanced proportionally to the number of emitters as a consequence of the formation of a collective superradiant state of the emitters \cite{dickeCoherenceSpontaneousRadiation1954, grossSuperradianceEssayTheory1982, trivediPhotonBlockadeWeakly2019,duanLongdistanceQuantumCommunication2001a, gonzalez-tudelaDeterministicGenerationArbitrary2015, paulischQuantumMetrologyOnedimensional2019}.

In practical devices, emitters do not have identical resonant frequencies \cite{evansNarrowLinewidthHomogeneousOptical2016, dibosAtomicSourceSingle2018a, zhongOpticallyAddressingSingle2018} — this inhomogeneous broadening in the resonant frequencies prohibits the formation of a collective superradiant state and lowers the transduction efficiencies. However, the temporal shape of the lasers driving the emitter ensembles can be experimentally tuned --- this opens up the possibility of using quantum control techniques to compensate for inhomogeneous broadening in the emitter ensemble, restore superradiance, and improve transduction efficiencies. 

Quantum control techniques have traditionally been employed to control the state of quantum systems \cite{dongQuantumControlTheory2010,kochControllingOpenQuantum2016} like ions \cite{grzesiakEfficientArbitrarySimultaneously2020, poulsenCorrectingErrorsQuantum2010}, atoms \cite{goerzRobustnessHighfidelityRydberg2014, treutleinMicrowavePotentialsOptimal2006, gorshkovPhotonStorageEnsuremath2008}, superconducting qubits \cite{goerzOptimalControlTheory2014, werninghausLeakageReductionFast2021, abdelhafezUniversalGatesProtected2020}, and solid-state emitters \cite{scheuerPreciseQubitControl2014, waldherrQuantumErrorCorrection2014}. Furthermore, several previous works have also applied quantum control techniques for addressing inhomogeneous ensembles for various quantum technology applications. However, most of these previous results consider an inhomogeneous non-interacting ensemble, in which case the system can be effectively analyzed with the density matrix of a single emitter obtained by averaging the individual inhomogeneous emitter trajectories. Several  results related to controllability of such systems have been previously provided \cite{liControlInhomogeneousQuantum2006, liEnsembleControlBloch2009, rabitzControllingQuantumDynamics2007, turiniciOptimallyControllingInternal2004}, together with analytical \cite{cumminsUseCompositeRotations2000, brownArbitrarilyAccurateComposite2004, owrutskyControlInhomogeneousEnsembles2012, anselSelectiveRobustTimeoptimal2021, augierAdiabaticEnsembleControl2018, tyckoBroadbandPopulationInversion1983, levittCompositePulses1986} and numerical techniques \cite{mischuckControlInhomogeneousAtomic2012, khaniHighfidelityQuantumGates2012, khanejaOptimalControlCoupled2005a, ruthsMultidimensionalPseudospectralMethod2011, chenSamplingbasedLearningControl2014, liOptimalPulseDesign2011, ruthsOptimalControlInhomogeneous2012, turiniciStochasticLearningControl2019, kuangRobustnessContinuousNonsmooth2020, wuLearningRobustHighprecision2019,turiniciOptimallyControllingInternal2004, sunEnsembleControlOpen2015, wangFreeendpointOptimalControl2018, kuangApproximateTimeoptimalControl2018, vandammeRobustOptimalControl2017,arjmandzadehQuantumGeneticLearning2017} to discover optimal controls. The problem of restoring superradiance in an inhomogeneous ensemble is distinct from the settings considered in these works in two key aspects --- first, we must necessarily account for the collective interaction between the different emitters mediated by the optical and microwave fields by considering the state of the \emph{entire} ensemble while designing the optimal control. Second, the model that we use is severely limited in terms of the control parameters available --- we do not assume that each emitter is individually accessible as practical experimental setups can only easily apply a single control signal across all the emitters. 

Our approach to solving this design problem is to use a time-dependent scattering theory framework \cite{trivediFewphotonScatteringEmission2018} to pose the problem of inhomogeneity compensation as a control problem --- this framework not only allows us to account for the collective interaction between the emitters as mediated by the optical and microwave fields, but also account for properties of the emitted and absorbed photons in the resulting quantum control problem. For the emitter based transduction system, we solve the resulting control problem using a gradient-based optimization algorithm to demonstrate an order of magnitude improvement in the transduction efficiencies. Furthermore, to assess the optimality of the resulting solution, we calculate provable upper bounds on the transduction efficiencies achievable by designing the temporal shape of the laser drive. Our work is closely related to, but distinct from Ref.~\cite{gorshkovPhotonStorageEnsuremath2008} wherein a similar framework was used to design quantum controls for mediating interactions between ensembles of emitters with controllable transition frequencies to implement quantum memories.

\section*{Results} \label{sec:results}

The transducer model being considered in this article is schematically depicted in Fig.~\ref{fig:1}a. The emitter ensemble, with each emitter considered to be a three-level system, is coupled to microwave and optical modes with coupling operators $L_{\mu}$ and $L_{\text{opt}}$ respectively, where
\begin{align*}
    L_{\mu} = \sum_{i = 1}^{N}\sqrt{\gamma_{\mu}}\sigma_{\mu}^{i} \ \text{and} \ L_{\text{opt}} = \sum_{i = 1}^{N} \sqrt{\gamma_{\text{opt}}} \sigma_{\text{opt}}^{ i }. \numberthis
\end{align*}


\begin{figure}[t]
    \centering
    \includegraphics[width=0.48\textwidth]{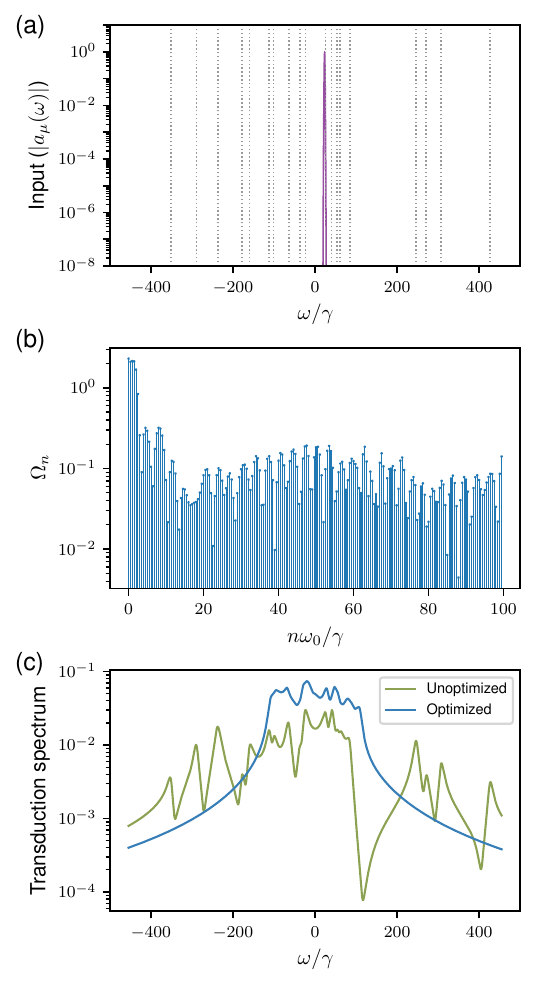}
    \caption{(a) Fourier transform of the input microwave field (Gaussian waveform). Dashed lines are representative of the individual emitter frequencies in a random ensemble ($N = 10,\  \Delta = 200 \gamma$). (b) Amplitudes of the harmonic components of the optimized $\Omega\left( t \right) $ designed for the same ensemble. (c) Comparison of the transduction spectrum of the same ensemble with and without optimized drives applied --- the transduction spectrum with the optimized drive is computed using a Floquet scattering theory approach \cite{trivediAnalyticGeometricProperties2020b}. }
    \label{fig:flo}
\end{figure}
Here, $\gamma_{\mu}$ and $\gamma_{\text{opt}}$ are the decay rates of the emitters into the microwave and optical modes respectively, $N$ is the number if emitters in the ensemble, and $\sigma_{\mu}^{i}$ and $ \sigma_{\text{opt}}^{ i  }$ are the lowering operators for transitions of the $i$th emitter in the ensemble. In addition to coupling to the optical and microwave modes, each emitter can also decay into additional loss channels, modeling unwanted radiative and non-radiative losses, with decay rates $\Gamma_{\mu}$ and $\Gamma_{\text{opt}}$ from the excited states $\ket{e_{\mu}^{ i  }}$ and $ \ket{e_{\text{opt}}^{ i  }}$, respectively. Furthermore, the transition between the two excited states is driven by a laser with envelope $\Omega\left( t \right) $.

For emitter ensembles formed out of identical emitters, the transduction efficiency is determined by the cooperativity of the individual transitions, $C_\mu = \gamma_\mu / \Gamma_\mu$ for microwave and $C_\text{opt} = \gamma_\text{opt} / \Gamma_\text{opt}$ for optical, as well as the number of emitters. We assume $\gamma_{\mu} = \gamma_{\text{opt}} = \gamma, \ \Gamma_{\mu} = \Gamma_{\text{opt}} = \Gamma$, and $C_{\mu} = C_{\text{opt}} = C = \gamma/\Gamma$ in our simulations for simplicity of analysis. Fig.~\ref{fig:1}b shows the transduction efficiency of this system as a function of the number of emitters for different emitter cooperativities — due to the formation of a collective superradiant state between the different emitters, this efficiency asymptotically reaches $1$ on increasing the number of emitters. Furthermore, the number of emitters needed to obtain high efficiency increases with a decrease in the cooperativity of the individual emitters. We point out that for high microwave and optical cooperativities, near unity transmissions can be obtained with a single emitter and consequently it is unnecessary to use emitter ensembles. We thus focus on low cooperativity emitters in the remainder of this article. On introducing inhomogeneous broadening into the emitter frequencies, the efficiency of the transduction system decreases (Fig.~\ref{fig:1}c) — for large inhomogeneous broadening, the emitters do not form a collective superradiant mode and the transduction spectrum simply comprises of the individual transduction spectra of the emitters in the ensemble (Fig.~\ref{fig:1}d). 

\begin{figure*}[t]
    \centering
    \includegraphics[width=\textwidth]{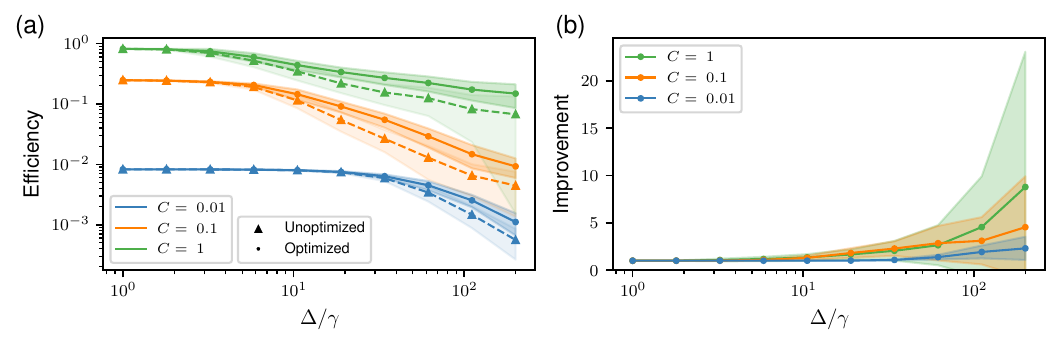}
    \caption{Optimized drives countering inhomogeneous broadening. (a) Transduction efficiency and (b) improvement in the transduction efficiency through randomly inhomogeneously broadened ensembles of $N = 10$ emitters with increasing inhomogeneous broadening for different cooperativities $C$ when the optimized drives are applied. For each $\Delta$, optimized drives are designed for each of the same 100 randomly generated ensembles with inhomogeneous broadening equal to $\Delta$ as used in Fig.~\ref{fig:1}c. Before running the optimizations, for each ensemble, the input photon is frequency-shifted to match the highest peak of the unoptimized transduction spectrum. Also, the initial condition for the optimization is $\Omega\left( t \right) = \left( N \gamma + \Gamma \right) / 2  $, which is a constant drive that maximises the transduction efficiency through a homogeneous ensemble with the same decay rates (see Appendix~\ref{sec:III}). Improvement is defined as the ratio of the efficiencies with and without the optimized drive applied. Each plot point corresponds to the mean over the 100 ensembles with inhomogeneous broadening equal to the corresponding value of $\Delta$ and the shaded regions represent the standard deviation.} 
    \label{fig:imp}
\end{figure*}
Since the laser pulse $\Omega(t)$ couples the microwave and optical transitions, we expect that unwanted variations in the transition frequencies can be compensated for by modulating the temporal form of this laser. However, in practical transduction systems, it is difficult to address individual emitters with separate lasers and consequently any modulation of $\Omega(t)$ impacts all the emitters. This makes designing the laser pulses difficult and calls for an application of numerical optimization techniques. We thus pose its design as maximizing the total power obtained in the optical mode when the emitter ensemble is excited with a single photon in the microwave mode:
\begin{align*}\label{eq:opt_problem}
    \max_{\Omega(t)} \quad & \int_{-\infty}^{ \infty} dt \ \abs{a_{\text{opt}}(t)}^2 \\
    \textrm{subject to} \quad & i \frac{d \ket{\psi_{\text{e}} \left( t \right)}}{dt} = H_{\text{eff}} \left( \Omega(t) \right) \ket{\psi_{\text{e}} \left( t \right)} + a_{\mu}(t) L_\mu^\dagger \ket{G},   \\
    \quad & a_{\text{opt}}(t) = -i\bra{G} L_\text{opt}\ket{\psi_{\text{e}} \left( t \right)}.\numberthis
\end{align*}
where the time-domain wave-packets of the single microwave input photon and optical output photon are described by $a_{\mu}\left( t \right) $ and $a_{\text{opt}}  \left( t \right) $ respectively, $\ket{\psi_{\text{e}}}$ is the state of the emitters in the ensemble,  $\ket{G}$ is the ground state of the ensemble, and $H_\text{eff}(\Omega)$ is the non-Hermitian effective Hamiltonian of the system when all the emitters are uniformly driven by a laser with amplitude $\Omega$. We point out that the  constraints are simply the input-output equations describing the dynamics of the transduction process under excitation with a single photon \cite{trivediFewphotonScatteringEmission2018, rephaeliFewPhotonSingleAtomCavity2012, fanInputoutputFormalismFewphoton2010} --- details of their derivation can be found in Appendix~\ref{sec:I}. Furthermore, since experimentally realizable laser pulses will be band-limited, we parametrize $\Omega\left( t \right) $ as a finite sum of harmonics,
\begin{align} \label{eq:parametrisation}
\Omega(t) = \sum_{n = 0}^{N_h} \Omega_n \cos(n\omega_0 t + \phi_n),
\end{align}
consequently constraining its bandwidth to be $N_h \omega_0$.
The design problem (\ref{eq:opt_problem}) can be solved using off-the-shelf gradient-based local optimizers. The gradient of the objective function in problem (\ref{eq:opt_problem}) with respect to the parameters  $\Omega_n, \phi_n $ can be computed using the time-domain adjoint variable method \cite{schmidtNumericalMethodsOptimal2006, swillamAccurateSensitivityAnalysis2007} (details available in Appendix~\ref{sec:V}).

As an example, we consider a transduction system with $N = 10$ inhomogeneous emitters excited with a single microwave photon with a Gaussian spectrum. Figure \ref{fig:flo}a shows the spectrum of the input photon, with the dashed lines depicting the resonant frequencies of the transduction spectra of the individual emitters. Given its narrow bandwidth, we expect the input photon to effectively only interact with a single emitter, leading to a low transduction efficiency comparable to what can be achieved by using just one emitter instead of many. The optimized drive obtained on solving problem (\ref{eq:opt_problem}) is depicted in Fig.~\ref{fig:flo}b --- as can be seen from Fig.~\ref{fig:flo}c, the transduction spectrum in the presence of the optimized drive shows improvement relative to the one with constant (unoptimized) drive.  

Statistical studies of performance of the optimization procedure for different sets of emitter frequencies is shown in Fig.~\ref{fig:imp} --- Fig.~\ref{fig:imp}a shows the optimized transduction efficiencies and Fig.~\ref{fig:imp}b shows the improvement in the transduction efficiencies. We observe that the improvements are larger at higher inhomogeneous broadening. Furthermore, the cooperativities of the emitters set a limit on improvement that can be obtained by shaping the laser pulse --- as can be seen from Fig.~\ref{fig:imp}b, the improvements are generally smaller for lower cooperativities. 
\begin{figure*}[t]
    \centering
    \includegraphics[width=\textwidth]{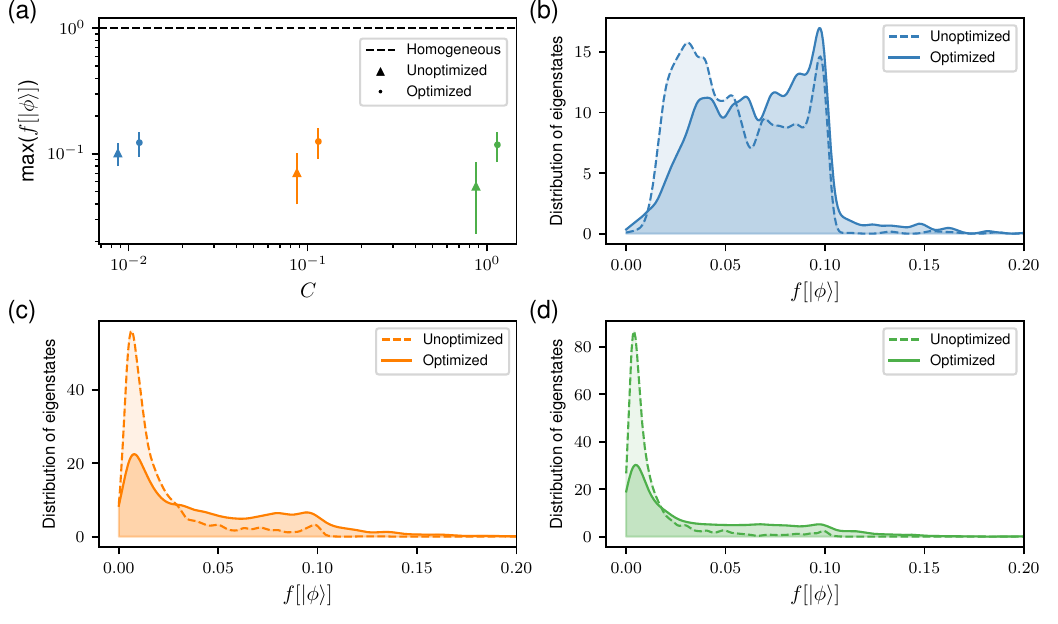}
    \caption{ (a) Comparison of the superradiance metric for ensembles with inhomogeneous broadening $\Delta = 200\gamma$ with and without optimized drives applied (data for optimized and unoptimized cases are dodged in the plot for visual clarity). After generating the optimized drives used in Fig.~\ref{fig:imp}, we compute the metric for all eigenstates of each of the 100 random ensembles with inhomogeneous broadening $\Delta = 200 \gamma$ by numerically diagonalising the propagator over one time period of the effective Hamiltonian. Each plot point and associated error bars correspond to the mean and standard deviation (over the collection of ensembles with $\Delta = 200 \gamma$) of the maximum value of the superradiance measure $f[\ket{\phi}]$ over all Floquet eigenstates $\ket{\phi} $. The dashed line denotes the same for a homogeneous ensemble. As we increase $\Gamma$ to decrease the cooperativity, the metric is larger on average in the unoptimized case. We attribute this to the simultaneous increase in the unoptimized drive $\Omega\left( t \right) = \left( N \gamma + \Gamma \right) / 2$ overshadowing the constant inhomogeneous broadening $\Delta = 200 \gamma$ (see Appendix~\ref{sec:IV}).  (b, c, d) Density plots (obtained by kernel density estimation using Gaussian kernels \cite{scottMultivariateDensityEstimation1992}) of the superradiance measure for eigenstates of the 100 ensembles with inhomogeneous broadening $\Delta = 200 \gamma$, (b) $C = 0.01$, (c)  $C = 0.1$, (d)  $C = 1$.     }
    \label{fig:metr}
\end{figure*}

While it is intuitively expected that improvement in transduction efficiency with the application of an optimized drive is due to recovery of superradiance, this can be made more concrete by studying the Floquet eigenstates of the optimized (time-dependent) effective Hamiltonian. The ‘superradiance’ in an eigenstate $\ket{\phi}$ of the propagator over one time period of the effective Hamiltonian, can be quantified with the metric,
\begin{align*} \label{eq:SRm}
    f[\ket{\phi}] = \frac{2}{N \sqrt{\gamma_{\mu} \gamma_{\text{opt}}}} \abs{\bra{G} L_{\text{opt}} \ket{\phi} \bra{\phi} L_{\mu}^{\dagger} \ket{G}}. \numberthis
\end{align*}
For a homogeneous ensemble, the metric is $1$ for two eigenstates formed by the drive-induced hybridization of superradiant states corresponding to the microwave and optical transitions. Furthermore, it is $0$ for the remaining eigenstates since they are subradiant/dark. Since the eigenstates for an inhomogeneous ensembles are not perfectly superradiant or subradiant, their corresponding metric lies between $0$ and $1$ and quantifies the extent of their subradiant or superradiant character. Figure~\ref{fig:metr}a indicates that an application of the optimized drive statistically increases the value of this metric, indicating partial recovery of superradiance. The density plots in Fig.~\ref{fig:metr}(b, c, d) show the distribution of the superradiance metric of the eigenstates of an inhomogeneously broadened ensemble. 

\begin{figure*}[t]
    \centering
    \includegraphics[width=\textwidth]{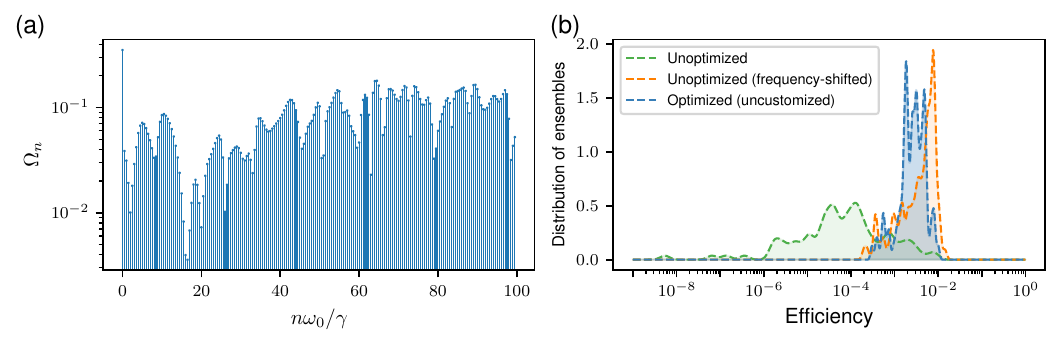}
    \caption{Transduction efficiency improvement with uncustomized optimization. (a) Amplitudes of the frequency components comprising the uncustomized drive. (b) Density plots of the transduction efficiency through 100 ensembles (test set) with $\Delta = 200 \gamma, \ C = 0.1$ for three cases -- (green) no optimised drive is applied and the input photon is fixed at the resonance of a homogeneous ensemble, (orange) no optimised drive is applied but the input photon is frequency-shifted to match the highest peak of the unoptimized transduction spectrum for each inhomogeneous ensemble, and (blue) the uncustomized optimized drive is applied and the input photon is fixed at the resonance of a homogeneous ensemble.}
    \label{fig:uncopt}
\end{figure*}

The results discussed above indicate that pulse-shaping the laser can be used to improve the performance of transduction systems. However, the optimized laser pulses can only be computed if the emitter frequencies are known. For systems with large number of emitters, such characterization might not be practical at scale and it would be desirable to find an optimized pulse which is robust to the specific frequencies of the emitters and depends only on their distribution. To design such a laser pulse, we modify the optimization problem (\ref{eq:opt_problem}) to
\begin{align*}\label{eq:opt_problem_unc}
    \max_{\Omega(t)} \quad & \frac{1}{N_s}\sum_{n = 1}^{N_s} \int_{-\infty}^{\infty} dt \ \abs{a^{(n)}_{\text{opt}}(t)}^2 \\
    \textrm{s.t.} \quad & i \frac{d \ket{\psi_{\text{e}}^{(n)} \left( t \right)}}{dt} = H_{\text{eff}}^{(n)} \left( \Omega(t) \right) \ket{\psi_{\text{e}}^{(n)} \left( t \right)} + a_{\mu}(t) L_\mu^\dagger \ket{G},   \\
    \quad & a^{(n)}_{\text{opt}}(t) = -i\bra{G} L_\text{opt}\ket{\psi_{\text{e}}^{(n)}\left( t \right)},\numberthis
\end{align*}
where we generate $N_s$ inhomogeneous emitter samples from the same inhomogeneous broadening distribution and find a laser pulse that $\Omega(t)$ that optimizes the average transduced power over all the samples. The superscript over a quantity in problem (\ref{eq:opt_problem_unc}) indicates that that quantity is computed for a specific sample. We design such a drive, shown in Fig.~\ref{fig:uncopt}a, for a training set of $N_s = 100$ random ensembles with inhomogeneous broadening $\Delta = 200 \gamma$ and with the input-photon being incident at the resonance of a homogeneous ensemble. Figure \ref{fig:uncopt}b shows the resulting improvement in transduction efficiency from applying the optimized drive to a test set of 100 random ensembles that are generated from the same inhomogeneous broadening distribution, independently of the training set. While there is significant improvement over the unoptimized case, we point out that simply shifting the spectrum of the input photon without shaping the driving laser pulse results in similar improvements. Therefore, it is not expected that this optimized drive is restoring superradiance in the emitter ensemble, rather it is effectively matching the resonance of the transduction spectrum to the input photon in a manner robust to the specific emitter frequencies. This could still be technologically useful since this optimized drive is agnostic to the specific emitter frequencies, thus obviating the need to characterize the emitter resonances. Furthermore, if many transducers are to be operated simultaneously, experimentally realizing and supplying drives customized to each transducer can be challenging to scale --- having a common, uncustomized drive would solve this problem. 

Finally, we address the question about the optimality of the laser pulses calculated using the gradient-based optimization algorithm. Since the optimization problem (\ref{eq:opt_problem}) is non-convex, we can only solve it locally and calculating the solution globally will likely be hard. However, one method to assess how close the laser pulses obtained above are to the globally optimal solution is to calculate upper bounds on the achievable transduction efficiency and compare it to the locally optimized results.

The physically motivated idea behind calculating such an upper bound is to note that the efficiency is limited by the amplitude of the emitters in their excited state while interacting with the input photon, as well as the time that the emitters spend in the excited state. More rigorously, in the presence of the incident single-photon wave-packet as well as a decay of the excited state, the time-integrated norm of the excited state amplitude $
\ket{\psi_e(t)}$ cannot be arbitrarily high. Consequently, an upper bound on the transduction efficiency can be obtained by simply maximizing the emitted photon energy as only constrained by this norm, which translates to solving the following optimization problem

\begin{align*}\label{eq:modified_opt_problem}
    \max_{\Omega(t)} \quad & \int_{- \infty}^{\infty} \abs{a_{\text{opt}}(t)}^2 \ dt\\
    \textrm{subject to} \quad & \int_{- \infty}^{ \infty} \Vert \ket{\psi_{\text{e}}(t)} - \ket{\psi_{\text{e}, 0}(t)} \Vert_2^2 \ dt \leq \varepsilon \\
    \quad & a_{\text{opt}}(t) = -i \bra{G} L_{\text{opt}} \ket{\psi_{\text{e}}(t)}, \numberthis
\end{align*}
where $\ket{\psi_{e, 0}(t)}$ is a reference state, $\Vert . \Vert_2$ denotes the $l_2$-\text{norm}, and $\varepsilon $ is parameter that can be considered as the solution of the following optimization problem:
\begin{align*} \label{eq:norm_problem}
    \max_{\Omega(t)} \quad & \int_{-\infty}^{\infty} \Vert \ket{\psi_{\text{e}}(t)} - \ket{\psi_{\text{e}, 0} (t)} \Vert_2^2 \ dt \\
    \textrm{subject to} \quad & i \frac{d \ket{\psi_{\text{e}} \left( t \right)}}{dt} = H_{\text{eff}} \left( \Omega(t) \right) \ket{\psi_{\text{e}} \left( t \right)} + a_{\mu}(t) L_\mu^\dagger \ket{G}. \numberthis
\end{align*}
We point out that since by construction $\varepsilon$ provides an upper bound on the integrated norm of the difference of the excited state from the reference state for all allowed laser pulses, the optimization problem \ref{eq:modified_opt_problem} is a \emph{relaxation} of the original non-convex optimization problem (problem \ref{eq:opt_problem}). Therefore, the solution of problem \ref{eq:modified_opt_problem} provides an upper bound to the (global) solution of problem \ref{eq:opt_problem}.

Problem (\ref{eq:modified_opt_problem}) is a quadratically-constrained quadratic program and bounds on its optimal value can be calculated by using the principle of Lagrangian duality \cite{boydConvexOptimization2004, trivediBoundsScatteringAbsorptionless2020} (see Appendix~\ref{sec:VI}). However, computing $\varepsilon$, which is required to solve problem \ref{eq:modified_opt_problem}, again requires solving a non-convex problem (problem \ref{eq:norm_problem}). In order to get around this issue, as outlined in appendix \ref{sec:VI}, we construct a provable upper bound, $\varepsilon_c$ on $\varepsilon$ which can also be used together with problem \ref{eq:modified_opt_problem} to obtain an upper bound on the transduction efficiency. We point out that this bound will be looser than the one obtained on using $\varepsilon$, i.e., the tighter the bound on the norm of the excited state, the better the bound on the transduction efficiency.

\begin{figure}[t]
    \centering
    \includegraphics[width=0.42\textwidth]{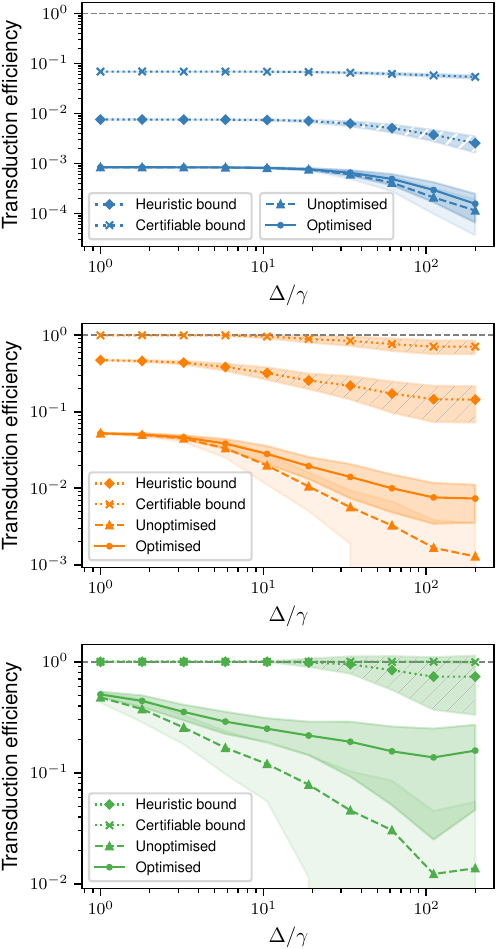}
    \caption{Heuristic and certifiable upper bounds and unoptimized and optimized transduction efficiencies calculated for ensembles with $N = 3$ emitters and cooperativities (a) $C = 0.01$, (b) $C = 0.1$, (c)  $C = 1$. For each $\Delta$, $100$ random ensembles are generated with inhomogeneous broadening equal to  $\Delta$. For each such ensemble, optimized drives are designed to improve transduction efficiency by using a local optimizer to solve problem (\ref{eq:opt_problem}). Then, using the state obtained by solving the input-output equation with the aforementioned optimized drive as the reference state, heuristic and certifiable bounds are calculated. Each plot point corresponds to the mean over the 100 ensembles with inhomogeneous broadening equal to the corresponding value of $\Delta$ and the shaded regions represent the standard deviation.}
    \label{fig:bounds}
\end{figure}

Fig.~\ref{fig:bounds} shows numerical studies of the upper bounds calculated on the transduction efficiency together with its comparison with the locally optimized results. In our numerical studies, we solve problem \ref{eq:modified_opt_problem} to compute both a \emph{certifiable} bound, which uses the upper bound $\varepsilon_c$ on $\varepsilon$, and a \emph{heuristic} bound calculated with only locally optimal solutions of problem \ref{eq:norm_problem}. We observe that, as physically expected, the bounds decrease on average with increasing inhomogeneous broadening and are higher for higher cooperativities. Furthermore, the optimized transduction efficiencies are within an order of magnitude of the bound, which provides us with an estimate of the performance of the optimization method used in the paper.

\section*{Discussion} 
\label{sec:disc}

In this article, we have used gradient-based inverse design of the temporal shape of the driving field as a technique to compensate for the effects of inhomogeneous broadening to help realize more efficient transducers. We demonstrated that optimized driving fields can lead to improvement in transduction efficiencies and showed that this improvement can be correlated with restoration of superradiant effects. Finally, to characterise the limits of the performance of time-dependent drives obtained by optimization-based design, we calculated upper bounds on optimal transduction efficiencies.

Our design method is applicable to different physical platforms including color centers or rare-earth ions in solid-state hosts. The techniques used in this article can be extended to ensembles that are orders of magnitude larger by frequency-binning the randomly distributed transition frequencies \cite{debnathCollectiveDynamicsInhomogeneously2019a}. We will explore this direction in future work. In some physical systems the transition frequencies of the emitters can be modulated (for e.g., via Stark effect in $\text{V}_{\text{Si}}$ centers in SiC). Previous research \cite{lukinSpectrallyReconfigurableQuantum2020} has shown that direct modulation of the transition frequencies can also be used to compensate for inhomogeneous broadening in a cavity-QED setting. We anticipate that optimization-based design for transducers can also be applied with the direct modulation as the degree of freedom instead of the driving field.

\section*{Methods} 
\label{sec:methods}

\subsection*{Simulations}
We discretize the input-output equations (the constraints in problem (\ref{eq:opt_problem})) in time and simulate the dynamics to calculate the transduction efficiency using finite-difference methods. For the customized case, i.e., when the drive is designed for a specific ensemble, we use the L-BFGS-B optimization algorithm. We employ the stochastic optimization algorithm \textit{Adam} \cite{kingmaAdamMethodStochastic2017} to design the uncustomized driving field. 

\section*{Acknowledgements}
The authors thank Shuo Sun, Logan Su, Hubert Stokowski, and Kevin Karan Singh Multani for useful discussions. This research is funded in part by the U.S. Department of Energy, Office of Science, under Awards DE-SC0019174 and DE-Ac02-76SF00515. R.T. acknowledges funding from Kailath Graduate Fellowship. 

S.D.M. and R.T. contributed equally to this work. R.T. and J.V. conceived the idea of using optimization-based design of drives for inhomogeneous broadening compensation. R.T., S.D.M., and A.H.S.-N. designed the numerical experiments. S.D.M. and R.T. performed the numerical and theoretical analysis. All authors wrote the manuscript.

\onecolumngrid
\newpage
\appendix
\section{Input-output equations}
\label{sec:I}

The Hamiltonian describing the ensemble is,
\begin{align*}\label{eq:Hsys}
    H_{\text{sys}}(\Omega(t)) = &\sum_i \left[ \delta_{\mu}^{i} \sigma_{\mu}^{i \dagger}\sigma_{\mu}^{i} + \delta_{\text{opt}}^{i} \sigma_{\text{opt}}^{i \dagger}\sigma_{\text{opt}}^{i} \right] + \sum_i \Omega (t) (\sigma_{\mu}^{i \dagger} \sigma_{\text{opt}}^{i} + \text{H. c.}), \numberthis
\end{align*} where the transition operators $\sigma_{\mu, \text{opt}} ^{ i }$ are defined in the main text. We point out that the laser field is actually $\Omega\left( t \right) e^{i \omega_L t}$, where $\omega_L$ is the central frequency. The Hamiltonian in Eq.~\ref{eq:Hsys} is obtained by going into a rotating frame to remove the term oscillating at $\omega_L$ from the drive.

The Hamiltonian of the entire system, i.e., the microwave and optical waveguide modes together with the ensemble is,
\begin{align*}\label{eq:H}
    H = -i \int dx \ \left(a_{\mu, x}^{\dagger} \frac{\partial}{\partial x} a_{\mu,x} + a_{\text{opt}, x}^{\dagger} \frac{\partial}{\partial x} a_{\text{opt},x}\right) + \left(a_{\mu, x = 0}^{\dagger} L_{\mu} + \text{H.c.}\right) + \left(a_{\text{opt}, x = 0}^{\dagger} L_{\text{opt}} + \text{H.c.}\right) + H_{\text{sys}}, \numberthis
\end{align*} where $a_{\mu, x}$ and $a_{\text{opt}, x}$ are the spatial annihilation operators for the microwave and optical waveguide modes respectively \cite{trivediFewphotonScatteringEmission2018} and the coupling operators $L_{\mu}$ and  $L_{\text{opt}}$ are defined in the main text. The terms in Eq.~\ref{eq:H} with the operators $L_{s}$ where $ s \in \{ \mu, \text{opt} \}$ represent the ensemble-waveguide interaction.

We define the number operator,
\begin{align*}\label{eq:Ne}
    N_e = \int dx \ \left( a_{\mu, x}^{\dagger} a_{\mu, x} + a_{\text{opt}, x}^{\dagger} a_{\text{opt}, x} \right) + \sum_{i = 1}^{N} \left( \sigma_{\mu}^{i \dagger} \sigma_{\mu}^{i}  + \sigma_{\text{opt}}^{i \dagger} \sigma_{\text{opt}}^{i} \right), \numberthis
\end{align*} which commutes with the Hamiltonian $H$. We consider an initial state with a single photon in the microwave waveguide mode. Thus, the state of the whole system is restricted to the single-excitation subspace at all times, and we assume the following ansatz for the state at time $t$ in the Schr\"{o}dinger picture, 
\begin{align*}\label{eq:ansatz}
\ket{\psi\left(t \right)} = \int dx \ \alpha\left( x, t \right) a_{\mu, x}^{\dagger} \ket{\text{vac}} \otimes \ket{G} \otimes \ket{\text{vac}} + \ket{\text{vac}} \otimes \ket{\psi_e\left(t  \right) } \otimes \ket{\text{vac}} + \ket{\text{vac}} \otimes \ket{G} \otimes \int dx \ \beta\left( x,t \right) a_{\text{opt},x}^{\dagger} \ket{\text{vac}}, \numberthis
\end{align*} where $\ket{\text{vac}} $ is the vacuum state of a waveguide mode and $\ket{G} = \bigotimes_{i=1}^{N} \ket{g}  $ is the ground state of the ensemble.

Given this ansatz, Schr\"{o}dinger's equations for the system are,
\begin{align}
&i \frac{d}{dt} \ket{\psi_e\left( t \right) } = H_{\text{sys}} \ket{\psi_{e}\left( t \right) } + \alpha\left( 0, t \right) L_{\mu}^{\dagger} \ket{G} + \beta\left( 0,t \right) L_{\text{opt}}^{\dagger}\ket{G} \label{eq:SE-sys} \\
& \frac{\partial}{\partial t} \alpha\left( x,t \right) = - \frac{\partial}{\partial x} \alpha\left( x,t \right) - i \delta\left( x \right) \bra{G} L_{\mu} \ket{\psi_e\left( t \right) } \label{eq:SE-mu} \\
& \frac{\partial}{\partial t} \beta\left( x,t \right) = - \frac{\partial}{\partial x} \beta\left( x,t \right) - i \delta\left( x \right) \bra{G} L_{\text{opt}} \ket{\psi_e\left( t \right) } \label{eq:SE-opt}, 
\end{align}

Solving Eq.~\ref{eq:SE-mu} and Eq.~\ref{eq:SE-opt} for $\alpha\left( 0,t \right) $ and $\beta\left( 0,t \right) $,
\begin{align}
    &\alpha(0,t) = a_{\mu}(t) - \frac{i}{2} \bra{G} L_{\mu} \ket{\psi_e(t)} \label{eq:alpha} \\
    &\beta(0,t) = - \frac{i}{2} \bra{G} L_{\text{opt}} \ket{\psi_e(t)}, \label{eq:beta}
\end{align} where $a_{\mu}\left( t \right) = \lim_{t_0 \to -\infty} \alpha \left( t_0 - t, t_0  \right)  $ describes the time-domain wave-packet of the input photon in the microwave waveguide mode. Similarly, $a_{\text{opt}}\left( t \right) = \lim_{t_1 \to \infty} \beta \left( t_1 - t, t_1  \right)  $ describes the time-domain wave-packet of the output photon in the optical waveguide mode. From the solution of Eq.~\ref{eq:SE-opt} we have, 
\begin{align} \label{eq:output}
&a_{\text{opt}} (t) = -i \bra{G} L_{\text{opt}} \ket{\psi_e(t)}.
\end{align}

Substituting Eq.~\ref{eq:alpha} and Eq.\ref{eq:beta} into Eq.~\ref{eq:SE-sys}, we have, 
\begin{align}
    i \frac{d}{dt} \ket{\psi_e\left( t \right) } = \left(H_{\text{sys}} - \frac{i}{2} L_{\mu}^{\dagger} \ket{G}\bra{G} L_{\mu} - \frac{i}{2} L_{\text{opt}}^{\dagger} \ket{G}\bra{G} L_{\text{opt}} \right) \ket{\psi_{e}\left( t \right) } + a_{\mu}(t) L_{\mu}^{\dagger} \ket{G}. \label{eq:dyn}
\end{align} We point out that $L_s^{\dagger} L_s = L_s^{\dagger} \left( P_e + \ket{G} \bra{G}  \right) L_s = L_s^{\dagger} \ket{G} \bra{G} L_s $ where $s \in \{ \mu, \text{opt} \}$ and $P_e$ is the projector onto the excited state space spanned by  $\{ \ket{e_{\mu}^{ i }}, \ket{e_{\text{opt}} ^{ i }} : i \in \{ 1, \ldots, N \}  \}$. Thus, Eq.~\ref{eq:dyn} can be rewritten as, 
\begin{align}
    i \frac{d}{dt} \ket{\psi_e\left( t \right) } = \left( H_{\text{sys}} - \frac{i}{2} L_{\mu}^{\dagger} L_{\mu} - \frac{i}{2} L_{\text{opt}}^{\dagger} L_{\text{opt}}\right) \ket{\psi_{e}\left( t \right) } + a_{\mu}(t) L_{\mu}^{\dagger} \ket{G}. \label{eq:iopre} 
\end{align}

Furthermore, the decays from the excited states of the emitters into various loss channels besides the waveguides can be captured in a similar manner by adding terms like $- \frac{i\Gamma_{s}^{i }}{2} \sigma_{s}^{ i \dagger } \sigma_{s}^{ i } \ket{\psi_e\left( t \right) }$ where $s \in \{\mu, \text{opt} \} $ to the right side of Eq.~\ref{eq:iopre}. Then we have,
\begin{align}
    i \frac{d}{dt} \ket{\psi_e\left( t \right) } = H_{\text{eff}}(\Omega(t)) \ket{\psi_{e}\left( t \right) } + a_{\mu}(t) L_{\mu}^{\dagger} \ket{G}, \label{eq:io} 
\end{align} where,
\begin{align}
    H_{\text{eff}}(\Omega(t)) = \left( H_{\text{sys}}(\Omega(t)) - \sum_i \left( \frac{i\Gamma_{\mu}}{2}\sigma_{\mu}^{i \dagger } \sigma_{\mu}^{ i } - \frac{i\Gamma_{\text{opt}}}{2}\sigma_{\text{opt}}^{i \dagger } \sigma_{\text{opt}}^{ i }   \right) - \frac{i}{2} L_{\mu}^{\dagger} L_{\mu} - \frac{i}{2} L_{\text{opt}}^{\dagger} L_{\text{opt}}\right).
\end{align}

Eq.~\ref{eq:io} and Eq.~\ref{eq:output} (upto a phase factor of $i$) are the input-output equations used in our simulations.

\section{Inhomogeneous broadening and transduction efficiency}
\label{sec:II}

We study how collective action of the emitters can be used to boost the effective cooperativity. For a homogeneous ensemble of size $N$, we define the operator-valued vectors, 

\begin{align*}
    \Sigma_{\mu} = \begin{pmatrix} \sigma_{\mu}^{ 1 } \\ \vdots \\ \sigma_{\mu}^ { N }  \end{pmatrix}, \Sigma_{\text{opt}} = \begin{pmatrix} \sigma_{\text{opt}}^{1 } \\ \vdots \\ \sigma_{\text{opt}}^ { N }  \end{pmatrix}
.\end{align*}
Next, we consider a change of basis through the action of a $N \times N$ unitary matrix $V$ on the vectors defined above. 
\begin{align*} \label{eq:5}
    S_{\mu} &= V \Sigma_{\mu} = \begin{pmatrix} S_{\mu, 1},&\ldots,&S_{\mu, N} \end{pmatrix}^T  \\
    S_{\text{opt}} &= V \Sigma_{\text{opt}} = \begin{pmatrix} S_{\text{opt}, 1},&\ldots,&S_{\text{opt}, N} \end{pmatrix}^T. \numberthis
\end{align*}
The transformation $V$ is defined such that 
\begin{align*}\label{eq:6}
    S_{\mu, 1} &= \frac{1}{\sqrt{N}} \sum_{i = 1}^{N} \sigma_{\mu}^{i} \\
    S_{\text{opt}, 1} &= \frac{1}{\sqrt{N}} \sum_{i = 1}^{N} \sigma_{\text{opt}}^{ i}. \numberthis 
\end{align*}
After the transformation, and assuming that $\gamma_{\mu} = \gamma_{\text{opt}} = \gamma, \ \Gamma_{\mu} = \Gamma_{\text{opt}} = \Gamma$, the effective Hamiltonian becomes, 
\begin{align*} \label{eq:Hhomo}
&H_{\text{eff}} = \left[ \left(\delta_{\mu} - \frac{i\Gamma}{2}\right) S_{\mu}^{\dagger}S_{\mu} + \left(\delta_{\text{opt}} - \frac{i \Gamma}{2} \right) S_{\text{opt}}^{ \dagger}S_{\text{opt}} \right] \\ & + \Omega (S_{\mu}^{\dagger} S_{\text{opt}}  + \text{H. c.}) -\frac{iN \gamma}{2}S_{\mu, 1}^{\dagger}S_{\mu, 1} - \frac{iN\gamma}   {2}S_{\text{opt}, 1}^{ \dagger}S_{\text{opt}, 1}, \numberthis
\end{align*} where $S_{s}^{\dagger} S_{s^{\prime}} \coloneqq \sum_{i = 1}^{N} S_{s, i}^{ \dagger} S_{s^{\prime}, i }$ for $s, s^{\prime} \in \{ \mu, \text{opt} \}$.
It can be seen from Eq.~\ref{eq:Hhomo} that the effective Hamiltonian is diagonal in the basis $\mathcal{B}$
\begin{align*}
\mathcal{B} = \{ S_{+, 1}^{ \dagger } \ket{G}, S_{-, 1}^{ \dagger } \ket{G}, \ldots, S_{+, N }^{ \dagger } \ket{G}, S_{-, N }^{ \dagger} \ket{G} \} \numberthis
\end{align*}
where $S_{+, i }^{ \dagger} = \left( S_{\mu, i }^{\dagger} + S_{\text{opt}, i }^{ \dagger} \right)/ \sqrt{2}$, and  $S_{-, i }^{ \dagger} = \left( S_{\mu, i }^{ \dagger} - S_{\text{opt},  i }^{ \dagger} \right)/ \sqrt{2}$. It is apparent that, in the diagonal basis, there are only two bright states $\{  S_{+,  1 }^{ \dagger } \ket{G}, S_{-,  1 }^{ \dagger } \ket{G}\}$; the other states don't couple to the waveguides due to orthogonality. The effective coupling rates of the bright states are enhanced by $N$ times to $N \gamma_{\mu}$ and $N\gamma_{\text{opt}}$, hence, the cooperativities are enhanced by $N$ times too. As the effective cooperativity scales linearly with the size of the ensemble, the maximum possible transduction efficiency should increase concurrently. For ensembles that are large enough, the maximum possible transduction efficiency should saturate to unity.

In the presence of inhomogeneous broadening, the advantage from scaling the number of emitters vanishes. The basis $\mathcal{B}$ identified earlier is no longer the diagonal basis for the effective Hamiltonian and the argument that we made in the case of a homogeneous ensemble that led to the scaling of the effective cooperativity can no longer be made. The collective action of the emitters, i.e., superradiance, is hampered by the differences in the emitters. In fact, with increasing inhomogeneous broadening, the maximum possible transmission drops to values corresponding to only a single emitter.

\section{Maximising transduction efficiency through a homogeneous ensemble}
\label{sec:III}

For a homogeneous ensemble, as there are only two bright states $\{  S_{+,  1 }^{ \dagger } \ket{G}, S_{-,  1 }^{\dagger } \ket{G}\}$ that couple to the waveguides, the ensemble can be equivalently considered to be a single emitter system with the coupling rates to the waveguides enhanced by $N$. The effective Hamiltonian of the equivalent single-emitter system is,
\begin{align}\label{eq:Heq}
H_{\text{eq}} =  \left(\delta_{\mu} - \frac{i(\Gamma + N \gamma) }{2}\right) S_{\mu, 1 }^{\dagger}S_{\mu, 1} + \left(\delta_{\text{opt}} - \frac{i (\Gamma + N \gamma)}{2} \right) S_{\text{opt}, 1 }^{\dagger}S_{\text{opt}, 1} + \Omega (S_{\mu, 1}^{ \dagger} S_{\text{opt},1}  + \text{H. c.}). 
\end{align} 

The transduction spectrum $\tau\left( \omega \right) $ through this system is given by \cite{trivediFewphotonScatteringEmission2018},
\begin{align}\label{eq:trspec}
\tau(\omega) = \sum_{s \in \{+,- \}} \frac{\bra{G} L_{\text{opt}} S_{s, 1}^{ \dagger} \ket{G} \bra{G} S_{s, 1} L_{\mu}^{\dagger} \ket{G}}{\omega - E_{s, 1}} 
\end{align} where $E_{\pm, 1   }$ are the eigenvalues of $H_{\text{eq}}$. Assuming $\delta_{\mu} =  \delta_{\text{opt}} = 0$, Eq.~\ref{eq:trspec} leads to,
\begin{align}
    | \tau(\omega) | = \frac{N \gamma \Omega}{|(\omega - E_{+,1}) (\omega - E_{-,1})|}
\end{align} where $E_{\pm,  i  } = \left( - \frac{i(\Gamma + N \gamma)}{2} \pm \Omega \right)$. For $ |\Omega| \ \leq \left( N \gamma + \Gamma \right) / 2 $, $ |\tau\left( \omega \right)| $ peaks at $\omega = 0$, with 
 \begin{align}
     |\tau(0)| = \frac{N \gamma \Omega}{\Omega^2 + \left(\frac{N \gamma + \Gamma}{2}\right)^2}.
 \end{align} 
 Hence, $|\tau\left( 0 \right)| $ is maximised when $\Omega = \left( N \gamma + \Gamma \right) / 2 $ and in that case, $|\tau\left( 0 \right)| = \frac{N \gamma}{\Gamma + N \gamma} $. We point out that when $|\Omega| \gg \left( N \gamma + \Gamma \right) / 2 $, the transduction spectrum has two peaks at $\approx \pm \Omega$, with $|\tau\left(\pm \Omega  \right) | = \frac{N \gamma}{\Gamma + N \gamma}$. As high drive strengths can be experimentally undesirable, we use the constant drive of least strength that maximises the transduction efficiency, i.e., $\Omega = \left( N \gamma + \Gamma\right) / 2 $, as the benchmark against which the optimized drives are judged.

\section{Hybridisation of eigenstates of a homogeneous ensemble with the introduction of inhomogeneity}
\label{sec:IV}

Consider the following separation of the effective Hamiltonian of an inhomogeneous ensemble into a Hamiltonian representing a homogeneous ensemble ($H_0$) and a perturbation term representing the inhomogeneity ($V$),
\begin{align}
    &H_{\text{eff}} = H_0 + \lambda V \\
    &H_0 = \sum_i \left[ -\frac{i \Gamma}{2} \sigma_{\mu}^{i \dagger}\sigma_{\mu}^{i} -\frac{i \Gamma}{2} \sigma_{\text{opt}}^{i \dagger}\sigma_{\text{opt}}^{i} \right] + \sum_i \Omega (\sigma_{\mu}^{i \dagger} \sigma_{\text{opt}}^{i} + \text{H. c.}) - \frac{i}{2} L_{\mu}^{\dagger}L_{\mu} - \frac{i}{2} L_{\text{opt}}^{\dagger}L_{\text{opt}}\\
    &V = \sum_i \left[ \delta_{\mu}^{i} \sigma_{\mu}^{i \dagger}\sigma_{\mu}^{i} + \delta_{\text{opt}}^{i} \sigma_{\text{opt}}^{i \dagger}\sigma_{\text{opt}}^{i} \right].
\end{align}

Defining, $\ket{E_{\pm,  i  }} = S_{\pm,  i }^{ \dagger} \ket{G} $, the eigenstates and eigenvalues of $H_0$ are,  

\begin{align}
&H_0 \ket{E_{\pm, 1}} = \left( - \frac{i(\Gamma + N \gamma)}{2} \pm \Omega \right) \ket{E_{\pm, 1}} \\
&H_0 \ket{E_{\pm, i}} = \left( - \frac{i \Gamma}{2} \pm \Omega \right) \ket{E_{\pm, i}}, \  i \neq 1 \\
\end{align}

With the introduction of the perturbation, the first order correction to the eigenstate $\ket{E_{+,\left( 1 \right) }} $ is,
\begin{align}\label{eq:pert}
    \ket{E_{+,1}} \rightarrow \ket{E_{+,i}} + \lambda \sum_{\substack{i \neq 1, \\ s \in \{+, -\}}} \ket{E_{s,i}} \frac{\bra{E_{s,i}} V \ket{E_{+,1}}}{E_{s,i} - E_{+,1}} + \lambda \ket{E_{-,1}} \frac{\bra{E_{-,1}} V \ket{E_{+,1}}}{E_{-,1} - E_{+,1}}.
\end{align}
The second term in Eq.~\ref{eq:pert} evaluates to zero and,
\begin{align}
    V \ket{E_{+, 1  }} = \frac{\left(\delta_{\mu}^{ 1 } + \delta_{\text{opt}}^{ 1  } \right) }{2} \ket{E_{+, 1  }} + \frac{\left(\delta_{\mu}^{ 1 } - \delta_{\text{opt}}^{ 1  } \right) }{2} \ket{E_{-, 1  }}.
\end{align} Thus, the perturbation changes the eigenstate as,
\begin{align}
    \ket{E_{+, 1}} \rightarrow \ket{E_{+, 1}} - \frac{\lambda (\delta_{\mu}^{1} - \delta_{\text{opt}}^{1})}{4 \Omega} \ket{E_{-,1}}.
\end{align} This indicates that the relevant frequency scale to compare the inhomogenous broadening to is the drive strength $\Omega$, which explains the variation of the superradiance metric with cooperativity in the unoptimized case in Fig. 4a. The unoptimized drive strength is chosen to be $\Omega(t) = \left( N \gamma + \Gamma \right) / 2 $ and as cooperativity is reduced by increasing $\Gamma$, the unoptimized drive strength increases, whereas the inhomogeneous broadening remains constant. Consequently, the hybridisation of the eigenstates away from that of a homogeneous ensemble decreases and the superradiance metric is higher on average. 

\section{Efficient calculation of gradients with the time-domain adjoint variable method}
\label{sec:V}

In this appendix, we describe how the adjoint variable method can be used to calculate the gradient with respect to all the parameters describing the laser drive efficiently in only two simulations (named forward and backward simulations). 

First, we rewrite Eq.~\ref{eq:io} and Eq.~\ref{eq:output} as,
\begin{align}
    & i \frac{dy(t)}{dt} = (H_0 + \Omega(t) H_1) y(t) + v_{\mu} a_{\mu}(t) \label{eq:io_re} \\
    & a_{\text{opt}}(t) = v_{opt}^{\dagger} y(t) \label{eq:output_re},
\end{align}
where $y\left( t \right) \coloneqq \ket{\psi\left( t \right) }$ is the vector describing the state of the ensemble, and
\begin{align}
    & H_0 \coloneqq \sum_i \left[ \delta_{\mu}^{i} \sigma_{\mu}^{i \dagger}\sigma_{\mu}^{i} + \delta_{\text{opt}}^{i} \sigma_{\text{opt}}^{i \dagger}\sigma_{\text{opt}}^{i} \right]  - \sum_i \left( \frac{i\Gamma_{\mu}}{2}\sigma_{\mu}^{i \dagger } \sigma_{\mu}^{i } - \frac{i\Gamma_{\text{opt}}}{2}\sigma_{\text{opt}}^{i \dagger } \sigma_{\text{opt}}^{i }   \right) - \frac{i}{2} L_{\mu}^{\dagger} L_{\mu} - \frac{i}{2} L_{\text{opt}}^{\dagger} L_{\text{opt}} \label{eq:H0} \\
    & H_1 \coloneqq  \sum_i (\sigma_{\mu}^{i \dagger} \sigma_{\text{opt}}^{i} + \text{H. c.}) \label{eq:H1} \\
    & v_{\mu} \coloneqq L_{\mu}^{\dagger} \ket{G}, \  v_{\text{opt}} \coloneqq i L_{\text{opt}}^{\dagger} \ket{G} \label{eq:conn_vectors}
\end{align}

We discretize the total simulation time range $[0, T]$ into $N - 1$ steps of duration $\delta t$ each. On this grid, the differential equation Eq.~\ref{eq:io_re} can be discretized as, 
\begin{align}
    & y[k + 1] = U[k] \left( y[k] - i \delta t \ a_{\mu}[k] \ v_{\mu}\right) \quad \quad \quad k \in \{0, \ldots, N - 2\} \label{eq:disc_ode} \\
    & y[0] = 0 \ (\text{initial condition}), \label{eq:init_cond}
\end{align}
where, $U[k] \coloneqq \text{exp}\left( -i \ \delta t \left( H_0 + \Omega[k] H_1 \right)  \right) $, $\Omega[k] \coloneqq \Omega\left( k \delta t \right) $, $y[k] \coloneqq y\left( k \delta t \right), \ a_{\mu}[k] \coloneqq a_{\mu}\left( k \delta t \right)  $. We refer to solving this system of equations as the forward simulation. 

The transduced power in the optical mode can be expressed in terms of the result of the forward simulation as,
\begin{align}
    P = \sum_{k = 0}^{N - 1}  |a_{\text{opt}}[k]|^2 = \sum_{k = 1}^{N-1}  |v_{\text{opt}}^{\dagger} y[k]|^2. 
\end{align} The latter sum starts from $k = 1$ as  $y[0] = 0$ is enforced as the initial condition.  

The derivative of $P$ with respect to the drive at the $l$th time step is, 
\begin{align} \label{eq:derivy}
     \frac{\partial P}{\partial \Omega[l]} = \sum_{k = 1}^{N - 1} r[k]^{\dagger} \frac{\partial y[k]}{\partial \Omega[l]} + C.c.,
\end{align}
where, `$C.c.$' means `complex conjugate' and $r[k] \coloneqq \left( v_{\text{opt}}^{\dagger} y[k]\right)^*  v_{\text{opt}}$.

We construct the following block vectors and matrices,
\begin{align}
    &y \coloneqq \begin{pmatrix} y[1] \\ y[2] \\ \vdots \\ y[N-1] \end{pmatrix}  \ r \coloneqq  \begin{pmatrix} r[1] \\ r[2] \\ \vdots \\ r[N-1] \end{pmatrix}  \ a \coloneqq -i \ \delta t \begin{pmatrix} a_{\mu}[1] v_{\mu} \\ a_{\mu}[2] v_{\mu} \\ \vdots \\ a_{\mu}[N-1] v_{\mu} \end{pmatrix} \\
    & M \coloneqq \begin{pmatrix} 0 &  & \cdots & & 0 \\
    U[1] & & &  & \\
     & U[2] &  &  & \vdots \\
    \vdots &  & \ddots &  &  \\
    0 & \cdots &  & U[N-2] & 0\end{pmatrix}  U \coloneqq \begin{pmatrix} U[0] & 0 & \cdots & 0 \\
    0 & U[1] & \cdots & 0 \\ 
    \vdots & \vdots & \ddots & \vdots \\
    0 & 0 & \cdots & U[N-2] \end{pmatrix}.
\end{align} The only non-zero blocks in the matrix $M$ are the subdiagonal blocks. 

The forward simulation Eq.~\ref{eq:disc_ode} can now be written as, 
\begin{align}
    y = My + Ua \label{eq:disc_complete}
\end{align}

As $\frac{\partial U[k]}{\partial \Omega[l]} = 0 \ \text{if} \  k \neq l$, taking the derivative of Eq.~\ref{eq:disc_complete}  with respect to the drive at the  $l$th time step results in, 
\begin{align}
    &\frac{\partial y}{\partial \Omega[l]} = M \frac{\partial y}{\partial \Omega[l]} + p[l] \\
    \Rightarrow &  \frac{\partial y}{\partial \Omega[l]} = (\mathds{1} - M)^{-1} p[l]  \quad \quad \quad l \in \{0, \ldots, N - 2\} \label{eq:derivstate}
\end{align}
where $\mathds{1}$ is the identity matrix and,
\begin{align}
    & p[l] = \begin{pmatrix} 0_{ld \times 1} \\ \alpha[l] \\ 0_{(N-l-2)d \times 1}  \end{pmatrix} \label{eq:p} \\
    & \alpha[l] = \frac{\partial U[l]}{\partial \Omega[l]}(y[l] + \delta t \  a_{\mu}[l] \  v_{\mu}).
\end{align}
In Eq.~\ref{eq:p}, $d$ is the dimension of the state vector  $y\left( t \right) $ and $0_{m \times n}$ is a zero matrix of dimension $m \times n$. To calculate $\alpha[l]$, we estimate the derivative $\frac{\partial U[l]}{\partial \Omega[l]}$ by a finite difference.

Using Eq.~\ref{eq:derivstate} in Eq.~\ref{eq:derivy}, we have, 
\begin{align}
    \frac{\partial P}{\partial \Omega[l]} = r^{\dagger}(\mathds{1} - M)^{-1}p[l] + C.c.
\end{align}
    We define a vector $q = \begin{pmatrix} q[1], & q[2], & \ldots, & q[N-1] \end{pmatrix}^T $ such that,  
\begin{align}
    & q^{\dagger} \coloneqq r^{\dagger}(\mathds{1} - M)^{-1} \\
    \Rightarrow & q = M^{\dagger} q + r \label {eq:adjoint}
\end{align}
Using the definition of $M$ and expanding out Eq.~\ref{eq:adjoint} in terms of the elements of $q$ and  $r$ results in the following system of equations,
\begin{align}
    &q[N-k] = U[N-k]^{\dagger}q[N-k+1] + r[N-k] \quad \quad \quad k \in \{2, 3, \ldots, N-1 \} \label{eq:adjoint_disc} \\
    &q[N-1] = r[N-1]. \label{eq:adjoint_initial}
\end{align}
We refer to $q[k]$ as the adjoint variables and to solving this system of equations (Eq.~\ref{eq:adjoint_disc} and Eq.~\ref{eq:adjoint_initial}) as the backward simulation. The initial condition for the backward simulation is provided at the final time point unlike the forward simulation. 

Once the forward and backward simulations are done, the gradient with respect to the drive at each time step can be computed just with an inner product of two $d-$dimensional vectors, 
\begin{align}
    & \frac{\partial P}{\partial \Omega[l]} = q^{\dagger}p[l] + C.c., \\
    \Rightarrow & \frac{\partial P}{\partial \Omega[l]} = q[l + 1]^{\dagger}\alpha[l] + C.c.  
\end{align}

The chain rule can then be used to compute the gradient with respect to the harmonic parameters $\Omega_n$ and  $\phi_n$,
\begin{align}
& \frac{\partial P}{\partial \Omega_n} = \sum_{l = 0}^{N - 2} \frac{\partial \Omega[l]}{\partial \Omega_n} \frac{\partial P}{\partial \Omega[l]} = \sum_{l = 0}^{N - 2} \cos{(n \omega_0 l \delta t + \phi_n)} \frac{\partial P}{\partial \Omega[l]}\\
    & \frac{\partial P}{\partial \phi_n} = \sum_{l = 0}^{N - 2} \frac{\partial \Omega[l]}{\partial \phi_n} \frac{\partial P}{\partial \Omega[l]} = \sum_{l = 0}^{N - 2} - \Omega_n \sin{(n \omega_0 l \delta t + \phi_n)} \frac{\partial P}{\partial \Omega[l]}
\end{align}

\section{Derivation of Lagrange duals of problems (\ref{eq:opt_problem}, \ref{eq:modified_opt_problem}), and upper bound on the optimal value of problem (\ref{eq:norm_problem})}
\label{sec:VI}

In this appendix, we show that the Lagrange dual for optimization problem (\ref{eq:opt_problem}) leads to a trivial bound, then we derive the Lagrange dual for the distance-constrained problem (\ref{eq:modified_opt_problem}), and finally provide a derivation for the upper bound $\varepsilon_c$ on the optimal value of problem (\ref{eq:norm_problem}).

For notational clarity, we rewrite problem (\ref{eq:opt_problem}) in the following manner,
\begin{align} 
    \max_{\Omega(t), y(t)} \quad & \int_{-\infty}^{\infty} y(t)^{\dagger} v_{\text{opt}} v_{\text{opt}}^{\dagger} y(t) \ dt \\
    \textrm{subject to} \quad &  \frac{dy(t)}{dt} = -i (H_0 + \Omega(t) H_1) y(t) -i v_{\mu} a_{\mu}(t),
\end{align}
where $y\left( t \right) \coloneqq \ket{\psi_{\text{e}}\left( t \right) }$ and $H_0, H_1, v_{\mu}, v_{\text{opt}}$ are defined in Eqs.~\ref{eq:H0}, \ref{eq:H1}, \ref{eq:conn_vectors}.

\subsection{Lagrange dual for problem (\ref{eq:opt_problem})}

Introducing the dual variables $\eta\left( t \right) $, the Lagrangian for problem (\ref{eq:opt_problem}) is:
\begin{align}
    \mathcal{L}(y(t), \Omega(t); \eta(t)) = \int_{-\infty}^{\infty} y(t)^{\dagger} V y(t) \ dt + 2 \int_{-\infty}^{\infty} \operatorname{Re} \left[ \eta(t)^{\dagger} \left( \frac{d y(t)}{dt} + iH_1 \Omega(t) y(t) + iH_0 y(t) + i v_{\mu} a_{\mu}(t) \right) \right],
\end{align}
where $\operatorname{Re}[.]$ denotes the real part of a complex number and $V \coloneqq v_{\text{opt}} v_{\text{opt}}^{\dagger}$. We point out that $V$ is a positive-definite matrix. 

Integrating by parts the term with the derivative of the state in the equation above, we have,
\begin{align}
    \mathcal{L}(y(t), \Omega(t); \eta(t)) = & \int_{-\infty}^{\infty} y(t)^{\dagger} V y(t) \ dt + \lim_{T \rightarrow \infty}  2 \operatorname{Re} \left[ \eta(T)^{\dagger} y(T) - \eta(-T)^{\dagger} y(-T)\right] + \\ & \int_{-\infty}^{\infty} 2 \operatorname{Re} \left[ z(t)^{\dagger}y(t) \right] \ dt + \int_{-\infty}^{\infty} 2 \operatorname{Re} \left[i \ \eta(t)^{\dagger} v_{\mu} a_{\mu}(t) \right],
\end{align}
where 
\begin{align}
    z(t) \coloneqq -\frac{d \eta(t)}{dt} - iH_1 \Omega(t) \eta(t) - iH_0 \eta(t). 
\end{align}
The Lagrange dual function $g\left( \eta\left( t \right)  \right) $ is given by, 
\begin{align}
    g(\eta(t)) \coloneqq \sup_{y(t), \Omega(t)} \mathcal{L}(y(t), \Omega(t); \eta(t)), 
\end{align} where $\sup{}$ denotes the supremum. As the matrix $V$ is positive-definite, the supremum of the Lagrangian over $y\left( t \right) $ is unbounded as the norm of $y\left( t \right) $ increases, and hence the dual function is unbounded too. Therefore, in this case, Lagrangian duality reveals a trivial upper bound on the transduced power. 

\subsection{Lagrange dual for problem (\ref{eq:modified_opt_problem})}

The Lagrangian for problem (\ref{eq:modified_opt_problem}) is 
 \begin{align}
     \mathcal{L}(y(t); \lambda) &= \int_{-\infty}^{\infty} y(t)^{\dagger} V y(t) \ dt + \lambda \left( \varepsilon - \int_{-\infty}^{\infty}\Vert y(t) - y_0(t) \Vert_2^2  \ dt \right) \\ 
                 &= \int_{-\infty}^{\infty} y(t)^{\dagger} \left(V - \lambda \mathds{1} \right) y(t) \ dt + \lambda \left( \varepsilon - \int_{-\infty}^{\infty} \Vert y_0(t) \Vert_2^2 \ dt \right) + \int_{-\infty}^{\infty} 2 \operatorname{Re} \left[ z(t)^{\dagger} y(t) \right] \ dt 
\end{align}
where $y_0\left( t \right) \coloneqq \ket{\psi_{\text{e},0}\left( t \right)} $ is the reference state, $\lambda$ is a dual variable and $z\left(t  \right) \coloneqq \lambda y_0\left( t \right) $. 
The corresponding dual function is, 
\begin{align}
    g(\lambda) &= \sup_{y(t)} \mathcal{L}(y(t); \lambda) \\
               &= \begin{cases} \int_{-\infty}^{\infty} z(t)^{\dagger} \left( \lambda \mathds{1} - V \right)^{-1} z(t) \ dt + \lambda \left( \varepsilon - \int_{-\infty}^{\infty} \Vert y_0(t) \Vert_2^2 \ dt \right), & \text{if} \ \lambda \mathds{1} - V \geq 0 \\ \infty, & \text{otherwise} \end{cases}
\end{align} 
 
To obtain the least upper bound, the dual function has to be minimised, leading to the following Lagrange dual problem, 
\begin{align*}
    \min_{\lambda, \beta(t)} \quad & \int_{-\infty}^{\infty} \beta(t) \ dt + \lambda \left( \varepsilon - \int_{-\infty}^{\infty} \Vert y_0(t) \Vert_2^2 \ dt \right)\\
    \textrm{subject to} \quad & \beta(t) \geq z(t)^{\dagger} (\lambda \mathds{1} - V)^{-1} z(t) \\
    \quad & \lambda \mathds{1} - V \geq 0,
\end{align*}
where we have introduced additional dual variables $\beta\left( t \right) $. 

\subsection{Upper bound on $\varepsilon^*$}

Eq.~\ref{eq:io_re} can further be rewritten as,
\begin{align}\label{eq:io_re_re}
    \frac{dy(t)}{dt} = (-i H_{\text{sys}}(\Omega(t)) - D) y(t) -i v_{\mu} a_{\mu}(t) 
\end{align}
where $H_{\text{sys}}\left(\Omega\left( t \right)   \right) $ is defined in Eq.~\ref{eq:Hsys} and,
\begin{align}
    D \coloneqq \sum_i \left( \frac{\Gamma_{\mu}}{2}\sigma_{\mu}^{ i  \dagger } \sigma_{\mu}^{ i  } + \frac{\Gamma_{\text{opt}}}{2}\sigma_{\text{opt}}^{ i  \dagger } \sigma_{\text{opt}}^{ i  }   \right) + \frac{1}{2} L_{\mu}^{\dagger} L_{\mu} + \frac{1}{2} L_{\text{opt}}^{\dagger} L_{\text{opt}}.
\end{align} 

Using the fact that $D^{\dagger} = D$, the time-evolution of the norm of the state can be written as,
\begin{align}
    \frac{dy(t)^{\dagger}y(t)}{dt} &= - 2 y(t)^{\dagger} D y(t) + 2 \operatorname{Re} \left[ -i y(t)^{\dagger} v_{\mu} a_{\mu}(t) \right] \\
                                   & \leq - 2 d_{\text{min}} y(t)^{\dagger} y(t) + 2 \operatorname{Re} \left[ -i y(t)^{\dagger} v_{\mu} a_{\mu}(t) \right], \label{eq:norm_ineq}
\end{align}
where $d_{\text{min}}$ is the smallest eigenvalue of $D$. 

From the inequality (\ref{eq:norm_ineq}) we have, 
\begin{align}
    \Vert y(t) \Vert_2^2 & \leq \int_0^t e^{-2 d_{\text{min}} (t-\tau)} 2 \operatorname{Re} \left[ -i y(\tau)^{\dagger} v_{\mu} a_{\mu}(\tau) \right] \ d \tau \\
                         & \leq \int_0^t 2 e^{-2 d_{\text{min}} (t-\tau)} \Vert y(\tau) \Vert_2 \Vert v_{\mu} a_{\mu}(\tau) \Vert_2 \ d \tau \label{eq:norm_ineq_line2} \\
                         & \leq \int_0^t 2 e^{-2 d_{\text{min}} (t-\tau)} \Vert v_{\mu} a_{\mu}(\tau) \Vert_2 \ d \tau \coloneqq d(t), \label{eq:norm_ineq_line3}
\end{align}
where, to go from (\ref{eq:norm_ineq_line2}) to (\ref{eq:norm_ineq_line3}) we use the fact $\Vert y\left( t \right) \Vert_2 \leq 1, \ \forall t$ . 

Therefore,
\begin{align}
    \int_{-\infty}^{\infty} \Vert y(t) - y_0(t) \Vert_2^2 \ dt & \leq \int_{-\infty}^{\infty} \left( \Vert y(t) \Vert_2 +\Vert y_0(t) \Vert_2 \right)^2 \ dt \\
                                                & \leq \int_{-\infty}^{\infty} \left( \sqrt{d(t)} + \Vert y_0(t) \Vert_2 \right)^2 \ dt \coloneqq \varepsilon_c
\end{align}

\section{Design of optimized drives with overlap-based objectives}
\label{sec:VII}

    Quantum information can be encoded in the temporal modes of photons for the purposes of quantum communication \cite{brechtPhotonTemporalModes2015}. Such encoding would necessitate a transduction process that preserves the fidelity of the transduced photon's wave-packet to specific temporal modes. We demonstrate in this appendix that it is possible to extend our design method to compensate for inhomogeneous broadening and produce improvements in transduction efficiency while preserving the overlap with a specified temporal mode. 

\begin{figure*}[t]
    \centering
    \includegraphics[width=\textwidth]{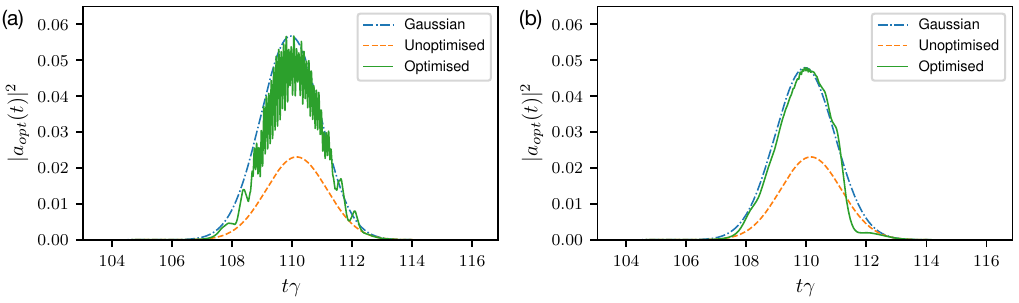}
    \caption{Temporal mode overlap-based design of drives. The amplitude $ \left|a_{\text{opt}} \left( t \right) \right|^2 $ of the output photon's temporal wave-packet after transduction by an ensemble of $N = 3$ emitters,  with $C = 0.1$, $\Delta \approx 61.61 \gamma$, and under the application of drives obtained by locally solving problem (\ref{eq:overlap_problem}) for (a) $c_0 \approx 1.331, c_1 = 0$ (b) $c_0 \approx 1.331, c_1 \approx 247.02$. The improvement in transduction efficiencies are $\left( a \right) 4.466 \times$ and $ \left( b \right) 3.977 \times $, respectively.}
    \label{fig:olap}
\end{figure*}

    To achieve this, we pose the design of the drive as maximising the overlap of the output photon's temporal wave-packet with a specified Hermite-Gaussian function \cite{courantSeriesExpansionsArbitrary1989} while simultaneously minimising its overlap with unwanted Hermite-Gaussian functions. For example,  
\begin{align}\label{eq:overlap_problem}
    \max_{\Omega(t)} \quad & \left| \int_{-\infty}^{\infty} dt \ c_0 \varphi_0(t) a_{\text{opt}}(t) \right|^2 -  \left| \int_{-\infty}^{\infty} dt \ c_1 \varphi_1(t) a_{\text{opt}}(t) \right|^2  \\
    \textrm{subject to} \quad & i \frac{d \ket{\psi_{\text{e}} \left( t \right)}}{dt} = H_{\text{eff}} \left( \Omega(t) \right) \ket{\psi_{\text{e}} \left( t \right)} + a_{\mu}(t) L_\mu^\dagger \ket{G},   \\
    \quad & a_{\text{opt}}(t) = -i\bra{G} L_\text{opt}\ket{\psi_{\text{e}} \left( t \right)}.\numberthis
\end{align}

where $\varphi_0\left(t \right) $ and $\varphi_1(t)$ are the 0th- and 1st-order Hermite-Gaussian functions (normalized to unity) \cite{courantSeriesExpansionsArbitrary1989} and the coefficients $c_0, c_1 \geq 0$ represent their weights. We solve problem (\ref{eq:overlap_problem}) for an input microwave photon with a Gaussian wave-packet incident on a randomly generated inhomogeneous ensemble with $N = 3$ emitters for two cases: $c_1 = 0$ and  $c_1 \neq 0$. Fig.~\ref{fig:olap} shows that, in both cases, we observe improved transduction efficiencies and the fidelity to the 0th-order mode is higher when $c_1 \neq 0$, in which case the output photon's shape closely resembles a Gaussian function. In case the input photon occupies a mode other than the 0th-order mode, this approach can be extended to design drives that improve the overlap of output photons with other Hermite-Gaussian modes too.


\bibliography{qtx}{}

\begin{thebibliography}{111}
\expandafter\ifx\csname natexlab\endcsname\relax\def\natexlab#1{#1}\fi
\expandafter\ifx\csname bibnamefont\endcsname\relax
  \def\bibnamefont#1{#1}\fi
\expandafter\ifx\csname bibfnamefont\endcsname\relax
  \def\bibfnamefont#1{#1}\fi
\expandafter\ifx\csname citenamefont\endcsname\relax
  \def\citenamefont#1{#1}\fi
\expandafter\ifx\csname url\endcsname\relax
  \def\url#1{\texttt{#1}}\fi
\expandafter\ifx\csname urlprefix\endcsname\relax\def\urlprefix{URL }\fi
\providecommand{\bibinfo}[2]{#2}
\providecommand{\eprint}[2][]{\url{#2}}

\bibitem[{\citenamefont{Arute et~al.}(2019)\citenamefont{Arute, Arya, Babbush,
  Bacon, Bardin, Barends, Biswas, Boixo, Brandao, Buell
  et~al.}}]{aruteQuantumSupremacyUsing2019}
\bibinfo{author}{\bibfnamefont{F.}~\bibnamefont{Arute}},
  \bibinfo{author}{\bibfnamefont{K.}~\bibnamefont{Arya}},
  \bibinfo{author}{\bibfnamefont{R.}~\bibnamefont{Babbush}},
  \bibinfo{author}{\bibfnamefont{D.}~\bibnamefont{Bacon}},
  \bibinfo{author}{\bibfnamefont{J.~C.} \bibnamefont{Bardin}},
  \bibinfo{author}{\bibfnamefont{R.}~\bibnamefont{Barends}},
  \bibinfo{author}{\bibfnamefont{R.}~\bibnamefont{Biswas}},
  \bibinfo{author}{\bibfnamefont{S.}~\bibnamefont{Boixo}},
  \bibinfo{author}{\bibfnamefont{F.~G. S.~L.} \bibnamefont{Brandao}},
  \bibinfo{author}{\bibfnamefont{D.~A.} \bibnamefont{Buell}},
  \bibnamefont{et~al.}, \bibinfo{journal}{Nature}
  \textbf{\bibinfo{volume}{574}}, \bibinfo{pages}{505} (\bibinfo{year}{2019}),
  ISSN \bibinfo{issn}{1476-4687}.

\bibitem[{\citenamefont{Kimble}(2008)}]{kimbleQuantumInternet2008a}
\bibinfo{author}{\bibfnamefont{H.~J.} \bibnamefont{Kimble}},
  \bibinfo{journal}{Nature} \textbf{\bibinfo{volume}{453}},
  \bibinfo{pages}{1023} (\bibinfo{year}{2008}), ISSN \bibinfo{issn}{1476-4687}.

\bibitem[{\citenamefont{Magnard et~al.}(2020)\citenamefont{Magnard, Storz,
  Kurpiers, Sch{\"a}r, Marxer, L{\"u}tolf, Besse, Gabureac, Reuer, Akin
  et~al.}}]{magnardMicrowaveQuantumLink2020}
\bibinfo{author}{\bibfnamefont{P.}~\bibnamefont{Magnard}},
  \bibinfo{author}{\bibfnamefont{S.}~\bibnamefont{Storz}},
  \bibinfo{author}{\bibfnamefont{P.}~\bibnamefont{Kurpiers}},
  \bibinfo{author}{\bibfnamefont{J.}~\bibnamefont{Sch{\"a}r}},
  \bibinfo{author}{\bibfnamefont{F.}~\bibnamefont{Marxer}},
  \bibinfo{author}{\bibfnamefont{J.}~\bibnamefont{L{\"u}tolf}},
  \bibinfo{author}{\bibfnamefont{J.-C.} \bibnamefont{Besse}},
  \bibinfo{author}{\bibfnamefont{M.}~\bibnamefont{Gabureac}},
  \bibinfo{author}{\bibfnamefont{K.}~\bibnamefont{Reuer}},
  \bibinfo{author}{\bibfnamefont{A.}~\bibnamefont{Akin}}, \bibnamefont{et~al.},
  \bibinfo{journal}{arXiv:2008.01642 [quant-ph]}  (\bibinfo{year}{2020}),
  \eprint{2008.01642}.

\bibitem[{\citenamefont{Lambert et~al.}(2020)\citenamefont{Lambert, Rueda,
  Sedlmeir, and Schwefel}}]{lambertCoherentConversionMicrowave2020}
\bibinfo{author}{\bibfnamefont{N.~J.} \bibnamefont{Lambert}},
  \bibinfo{author}{\bibfnamefont{A.}~\bibnamefont{Rueda}},
  \bibinfo{author}{\bibfnamefont{F.}~\bibnamefont{Sedlmeir}}, \bibnamefont{and}
  \bibinfo{author}{\bibfnamefont{H.~G.~L.} \bibnamefont{Schwefel}},
  \bibinfo{journal}{Advanced Quantum Technologies}
  \textbf{\bibinfo{volume}{3}}, \bibinfo{pages}{1900077}
  (\bibinfo{year}{2020}), ISSN \bibinfo{issn}{2511-9044}.

\bibitem[{\citenamefont{Lauk et~al.}(2020)\citenamefont{Lauk, Sinclair,
  Barzanjeh, Covey, Saffman, Spiropulu, and
  Simon}}]{laukPerspectivesQuantumTransduction2020}
\bibinfo{author}{\bibfnamefont{N.}~\bibnamefont{Lauk}},
  \bibinfo{author}{\bibfnamefont{N.}~\bibnamefont{Sinclair}},
  \bibinfo{author}{\bibfnamefont{S.}~\bibnamefont{Barzanjeh}},
  \bibinfo{author}{\bibfnamefont{J.~P.} \bibnamefont{Covey}},
  \bibinfo{author}{\bibfnamefont{M.}~\bibnamefont{Saffman}},
  \bibinfo{author}{\bibfnamefont{M.}~\bibnamefont{Spiropulu}},
  \bibnamefont{and} \bibinfo{author}{\bibfnamefont{C.}~\bibnamefont{Simon}},
  \bibinfo{journal}{Quantum Science and Technology}
  \textbf{\bibinfo{volume}{5}}, \bibinfo{pages}{020501} (\bibinfo{year}{2020}),
  ISSN \bibinfo{issn}{2058-9565}.

\bibitem[{\citenamefont{McKenna et~al.}(2020)\citenamefont{McKenna, Witmer,
  Patel, Jiang, Van~Laer, {Arrangoiz-Arriola}, Wollack, Herrmann, and
  {Safavi-Naeini}}}]{mckennaCryogenicMicrowavetoopticalConversion2020}
\bibinfo{author}{\bibfnamefont{T.~P.} \bibnamefont{McKenna}},
  \bibinfo{author}{\bibfnamefont{J.~D.} \bibnamefont{Witmer}},
  \bibinfo{author}{\bibfnamefont{R.~N.} \bibnamefont{Patel}},
  \bibinfo{author}{\bibfnamefont{W.}~\bibnamefont{Jiang}},
  \bibinfo{author}{\bibfnamefont{R.}~\bibnamefont{Van~Laer}},
  \bibinfo{author}{\bibfnamefont{P.}~\bibnamefont{{Arrangoiz-Arriola}}},
  \bibinfo{author}{\bibfnamefont{E.~A.} \bibnamefont{Wollack}},
  \bibinfo{author}{\bibfnamefont{J.~F.} \bibnamefont{Herrmann}},
  \bibnamefont{and} \bibinfo{author}{\bibfnamefont{A.~H.}
  \bibnamefont{{Safavi-Naeini}}}, \bibinfo{journal}{arXiv:2005.00897 [physics,
  physics:quant-ph]}  (\bibinfo{year}{2020}), \bibinfo{note}{comment: 15 pages,
  10 figures. First two authors contributed equally to this work},
  \eprint{2005.00897}.

\bibitem[{\citenamefont{Holzgrafe et~al.}(2020)\citenamefont{Holzgrafe,
  Sinclair, Zhu, {Shams-Ansari}, Colangelo, Hu, Zhang, Berggren, and Lon{\v
  c}ar}}]{holzgrafeCavityElectroopticsThinfilm2020}
\bibinfo{author}{\bibfnamefont{J.}~\bibnamefont{Holzgrafe}},
  \bibinfo{author}{\bibfnamefont{N.}~\bibnamefont{Sinclair}},
  \bibinfo{author}{\bibfnamefont{D.}~\bibnamefont{Zhu}},
  \bibinfo{author}{\bibfnamefont{A.}~\bibnamefont{{Shams-Ansari}}},
  \bibinfo{author}{\bibfnamefont{M.}~\bibnamefont{Colangelo}},
  \bibinfo{author}{\bibfnamefont{Y.}~\bibnamefont{Hu}},
  \bibinfo{author}{\bibfnamefont{M.}~\bibnamefont{Zhang}},
  \bibinfo{author}{\bibfnamefont{K.~K.} \bibnamefont{Berggren}},
  \bibnamefont{and} \bibinfo{author}{\bibfnamefont{M.}~\bibnamefont{Lon{\v
  c}ar}}, \bibinfo{journal}{arXiv:2005.00939 [physics, physics:quant-ph]}
  (\bibinfo{year}{2020}), \eprint{2005.00939}.

\bibitem[{\citenamefont{Soltani et~al.}(2017)\citenamefont{Soltani, Zhang,
  Ryan, Ribeill, Wang, and
  Loncar}}]{soltaniEfficientQuantumMicrowavetooptical2017a}
\bibinfo{author}{\bibfnamefont{M.}~\bibnamefont{Soltani}},
  \bibinfo{author}{\bibfnamefont{M.}~\bibnamefont{Zhang}},
  \bibinfo{author}{\bibfnamefont{C.}~\bibnamefont{Ryan}},
  \bibinfo{author}{\bibfnamefont{G.~J.} \bibnamefont{Ribeill}},
  \bibinfo{author}{\bibfnamefont{C.}~\bibnamefont{Wang}}, \bibnamefont{and}
  \bibinfo{author}{\bibfnamefont{M.}~\bibnamefont{Loncar}},
  \bibinfo{journal}{Physical Review A} \textbf{\bibinfo{volume}{96}},
  \bibinfo{pages}{043808} (\bibinfo{year}{2017}).

\bibitem[{\citenamefont{Tsang}(2010)}]{tsangCavityQuantumElectrooptics2010a}
\bibinfo{author}{\bibfnamefont{M.}~\bibnamefont{Tsang}},
  \bibinfo{journal}{Physical Review A} \textbf{\bibinfo{volume}{81}},
  \bibinfo{pages}{063837} (\bibinfo{year}{2010}).

\bibitem[{\citenamefont{Rueda et~al.}(2019)\citenamefont{Rueda, Hease,
  Barzanjeh, and Fink}}]{ruedaElectroopticEntanglementSource2019}
\bibinfo{author}{\bibfnamefont{A.}~\bibnamefont{Rueda}},
  \bibinfo{author}{\bibfnamefont{W.}~\bibnamefont{Hease}},
  \bibinfo{author}{\bibfnamefont{S.}~\bibnamefont{Barzanjeh}},
  \bibnamefont{and} \bibinfo{author}{\bibfnamefont{J.~M.} \bibnamefont{Fink}},
  \bibinfo{journal}{npj Quantum Information} \textbf{\bibinfo{volume}{5}},
  \bibinfo{pages}{1} (\bibinfo{year}{2019}), ISSN \bibinfo{issn}{2056-6387}.

\bibitem[{\citenamefont{Rueda et~al.}(2016)\citenamefont{Rueda, Sedlmeir,
  Collodo, Vogl, Stiller, Schunk, Strekalov, Marquardt, Fink, Painter
  et~al.}}]{ruedaEfficientMicrowaveOptical2016}
\bibinfo{author}{\bibfnamefont{A.}~\bibnamefont{Rueda}},
  \bibinfo{author}{\bibfnamefont{F.}~\bibnamefont{Sedlmeir}},
  \bibinfo{author}{\bibfnamefont{M.~C.} \bibnamefont{Collodo}},
  \bibinfo{author}{\bibfnamefont{U.}~\bibnamefont{Vogl}},
  \bibinfo{author}{\bibfnamefont{B.}~\bibnamefont{Stiller}},
  \bibinfo{author}{\bibfnamefont{G.}~\bibnamefont{Schunk}},
  \bibinfo{author}{\bibfnamefont{D.~V.} \bibnamefont{Strekalov}},
  \bibinfo{author}{\bibfnamefont{C.}~\bibnamefont{Marquardt}},
  \bibinfo{author}{\bibfnamefont{J.~M.} \bibnamefont{Fink}},
  \bibinfo{author}{\bibfnamefont{O.}~\bibnamefont{Painter}},
  \bibnamefont{et~al.}, \bibinfo{journal}{Optica} \textbf{\bibinfo{volume}{3}},
  \bibinfo{pages}{597} (\bibinfo{year}{2016}), ISSN \bibinfo{issn}{2334-2536}.

\bibitem[{\citenamefont{Hisatomi et~al.}(2016)\citenamefont{Hisatomi, Osada,
  Tabuchi, Ishikawa, Noguchi, Yamazaki, Usami, and
  Nakamura}}]{hisatomiBidirectionalConversionMicrowave2016a}
\bibinfo{author}{\bibfnamefont{R.}~\bibnamefont{Hisatomi}},
  \bibinfo{author}{\bibfnamefont{A.}~\bibnamefont{Osada}},
  \bibinfo{author}{\bibfnamefont{Y.}~\bibnamefont{Tabuchi}},
  \bibinfo{author}{\bibfnamefont{T.}~\bibnamefont{Ishikawa}},
  \bibinfo{author}{\bibfnamefont{A.}~\bibnamefont{Noguchi}},
  \bibinfo{author}{\bibfnamefont{R.}~\bibnamefont{Yamazaki}},
  \bibinfo{author}{\bibfnamefont{K.}~\bibnamefont{Usami}}, \bibnamefont{and}
  \bibinfo{author}{\bibfnamefont{Y.}~\bibnamefont{Nakamura}},
  \bibinfo{journal}{Physical Review B} \textbf{\bibinfo{volume}{93}},
  \bibinfo{pages}{174427} (\bibinfo{year}{2016}).

\bibitem[{\citenamefont{Everts et~al.}(2019)\citenamefont{Everts, Berrington,
  Ahlefeldt, and Longdell}}]{evertsMicrowaveOpticalPhoton2019a}
\bibinfo{author}{\bibfnamefont{J.~R.} \bibnamefont{Everts}},
  \bibinfo{author}{\bibfnamefont{M.~C.} \bibnamefont{Berrington}},
  \bibinfo{author}{\bibfnamefont{R.~L.} \bibnamefont{Ahlefeldt}},
  \bibnamefont{and} \bibinfo{author}{\bibfnamefont{J.~J.}
  \bibnamefont{Longdell}}, \bibinfo{journal}{Physical Review A}
  \textbf{\bibinfo{volume}{99}}, \bibinfo{pages}{063830}
  (\bibinfo{year}{2019}).

\bibitem[{\citenamefont{Everts et~al.}(2020)\citenamefont{Everts, King,
  Lambert, Kocsis, Rogge, and
  Longdell}}]{evertsUltrastrongCouplingMicrowave2020}
\bibinfo{author}{\bibfnamefont{J.}~\bibnamefont{Everts}},
  \bibinfo{author}{\bibfnamefont{G.~G.~G.} \bibnamefont{King}},
  \bibinfo{author}{\bibfnamefont{N.}~\bibnamefont{Lambert}},
  \bibinfo{author}{\bibfnamefont{S.}~\bibnamefont{Kocsis}},
  \bibinfo{author}{\bibfnamefont{S.}~\bibnamefont{Rogge}}, \bibnamefont{and}
  \bibinfo{author}{\bibfnamefont{J.~J.} \bibnamefont{Longdell}},
  \bibinfo{journal}{Physical Review B} \textbf{\bibinfo{volume}{101}},
  \bibinfo{pages}{214414} (\bibinfo{year}{2020}), ISSN
  \bibinfo{issn}{2469-9950, 2469-9969}, \bibinfo{note}{comment: 5 pages, 5
  figures}, \eprint{1911.11311}.

\bibitem[{\citenamefont{Zhong et~al.}(2020)\citenamefont{Zhong, Wang, Zou,
  Zhang, Han, Fu, Xu, Shankar, Devoret, Tang
  et~al.}}]{zhongProposalHeraldedGeneration2020}
\bibinfo{author}{\bibfnamefont{C.}~\bibnamefont{Zhong}},
  \bibinfo{author}{\bibfnamefont{Z.}~\bibnamefont{Wang}},
  \bibinfo{author}{\bibfnamefont{C.}~\bibnamefont{Zou}},
  \bibinfo{author}{\bibfnamefont{M.}~\bibnamefont{Zhang}},
  \bibinfo{author}{\bibfnamefont{X.}~\bibnamefont{Han}},
  \bibinfo{author}{\bibfnamefont{W.}~\bibnamefont{Fu}},
  \bibinfo{author}{\bibfnamefont{M.}~\bibnamefont{Xu}},
  \bibinfo{author}{\bibfnamefont{S.}~\bibnamefont{Shankar}},
  \bibinfo{author}{\bibfnamefont{M.~H.} \bibnamefont{Devoret}},
  \bibinfo{author}{\bibfnamefont{H.~X.} \bibnamefont{Tang}},
  \bibnamefont{et~al.}, \bibinfo{journal}{Physical Review Letters}
  \textbf{\bibinfo{volume}{124}}, \bibinfo{pages}{010511}
  (\bibinfo{year}{2020}).

\bibitem[{\citenamefont{Wu et~al.}(2020)\citenamefont{Wu, Zeuthen, Balram, and
  Srinivasan}}]{wuMicrowavetoOpticalTransductionUsing2020}
\bibinfo{author}{\bibfnamefont{M.}~\bibnamefont{Wu}},
  \bibinfo{author}{\bibfnamefont{E.}~\bibnamefont{Zeuthen}},
  \bibinfo{author}{\bibfnamefont{K.~C.} \bibnamefont{Balram}},
  \bibnamefont{and}
  \bibinfo{author}{\bibfnamefont{K.}~\bibnamefont{Srinivasan}},
  \bibinfo{journal}{Physical Review Applied} \textbf{\bibinfo{volume}{13}},
  \bibinfo{pages}{014027} (\bibinfo{year}{2020}).

\bibitem[{\citenamefont{Lau and Clerk}(2020)}]{lauGroundStateCooling2020}
\bibinfo{author}{\bibfnamefont{H.-K.} \bibnamefont{Lau}} \bibnamefont{and}
  \bibinfo{author}{\bibfnamefont{A.~A.} \bibnamefont{Clerk}},
  \bibinfo{journal}{Physical Review Letters} \textbf{\bibinfo{volume}{124}},
  \bibinfo{pages}{103602} (\bibinfo{year}{2020}), ISSN
  \bibinfo{issn}{0031-9007, 1079-7114}, \bibinfo{note}{comment: Close to
  accepted version}, \eprint{1904.12984}.

\bibitem[{\citenamefont{Jiang et~al.}(2020)\citenamefont{Jiang, Sarabalis,
  Dahmani, Patel, Mayor, McKenna, Laer, and
  {Safavi-Naeini}}}]{jiangEfficientBidirectionalPiezooptomechanical2020}
\bibinfo{author}{\bibfnamefont{W.}~\bibnamefont{Jiang}},
  \bibinfo{author}{\bibfnamefont{C.~J.} \bibnamefont{Sarabalis}},
  \bibinfo{author}{\bibfnamefont{Y.~D.} \bibnamefont{Dahmani}},
  \bibinfo{author}{\bibfnamefont{R.~N.} \bibnamefont{Patel}},
  \bibinfo{author}{\bibfnamefont{F.~M.} \bibnamefont{Mayor}},
  \bibinfo{author}{\bibfnamefont{T.~P.} \bibnamefont{McKenna}},
  \bibinfo{author}{\bibfnamefont{R.~V.} \bibnamefont{Laer}}, \bibnamefont{and}
  \bibinfo{author}{\bibfnamefont{A.~H.} \bibnamefont{{Safavi-Naeini}}},
  \bibinfo{journal}{Nature Communications} \textbf{\bibinfo{volume}{11}},
  \bibinfo{pages}{1} (\bibinfo{year}{2020}), ISSN \bibinfo{issn}{2041-1723}.

\bibitem[{\citenamefont{Forsch et~al.}(2020)\citenamefont{Forsch, Stockill,
  Wallucks, Marinkovi{\'c}, G{\"a}rtner, Norte, van Otten, Fiore, Srinivasan,
  and Gr{\"o}blacher}}]{forschMicrowavetoopticsConversionUsing2020}
\bibinfo{author}{\bibfnamefont{M.}~\bibnamefont{Forsch}},
  \bibinfo{author}{\bibfnamefont{R.}~\bibnamefont{Stockill}},
  \bibinfo{author}{\bibfnamefont{A.}~\bibnamefont{Wallucks}},
  \bibinfo{author}{\bibfnamefont{I.}~\bibnamefont{Marinkovi{\'c}}},
  \bibinfo{author}{\bibfnamefont{C.}~\bibnamefont{G{\"a}rtner}},
  \bibinfo{author}{\bibfnamefont{R.~A.} \bibnamefont{Norte}},
  \bibinfo{author}{\bibfnamefont{F.}~\bibnamefont{van Otten}},
  \bibinfo{author}{\bibfnamefont{A.}~\bibnamefont{Fiore}},
  \bibinfo{author}{\bibfnamefont{K.}~\bibnamefont{Srinivasan}},
  \bibnamefont{and}
  \bibinfo{author}{\bibfnamefont{S.}~\bibnamefont{Gr{\"o}blacher}},
  \bibinfo{journal}{Nature Physics} \textbf{\bibinfo{volume}{16}},
  \bibinfo{pages}{69} (\bibinfo{year}{2020}), ISSN \bibinfo{issn}{1745-2481}.

\bibitem[{\citenamefont{Arnold et~al.}(2020)\citenamefont{Arnold, Wulf,
  Barzanjeh, Redchenko, Rueda, Hease, Hassani, and
  Fink}}]{arnoldConvertingMicrowaveTelecom2020}
\bibinfo{author}{\bibfnamefont{G.}~\bibnamefont{Arnold}},
  \bibinfo{author}{\bibfnamefont{M.}~\bibnamefont{Wulf}},
  \bibinfo{author}{\bibfnamefont{S.}~\bibnamefont{Barzanjeh}},
  \bibinfo{author}{\bibfnamefont{E.~S.} \bibnamefont{Redchenko}},
  \bibinfo{author}{\bibfnamefont{A.}~\bibnamefont{Rueda}},
  \bibinfo{author}{\bibfnamefont{W.~J.} \bibnamefont{Hease}},
  \bibinfo{author}{\bibfnamefont{F.}~\bibnamefont{Hassani}}, \bibnamefont{and}
  \bibinfo{author}{\bibfnamefont{J.~M.} \bibnamefont{Fink}},
  \bibinfo{journal}{Nature Communications} \textbf{\bibinfo{volume}{11}},
  \bibinfo{pages}{4460} (\bibinfo{year}{2020}), ISSN \bibinfo{issn}{2041-1723}.

\bibitem[{\citenamefont{Bagci et~al.}(2014)\citenamefont{Bagci, Simonsen,
  Schmid, Villanueva, Zeuthen, Appel, Taylor, S{\o}rensen, Usami, Schliesser
  et~al.}}]{bagciOpticalDetectionRadio2014a}
\bibinfo{author}{\bibfnamefont{T.}~\bibnamefont{Bagci}},
  \bibinfo{author}{\bibfnamefont{A.}~\bibnamefont{Simonsen}},
  \bibinfo{author}{\bibfnamefont{S.}~\bibnamefont{Schmid}},
  \bibinfo{author}{\bibfnamefont{L.~G.} \bibnamefont{Villanueva}},
  \bibinfo{author}{\bibfnamefont{E.}~\bibnamefont{Zeuthen}},
  \bibinfo{author}{\bibfnamefont{J.}~\bibnamefont{Appel}},
  \bibinfo{author}{\bibfnamefont{J.~M.} \bibnamefont{Taylor}},
  \bibinfo{author}{\bibfnamefont{A.}~\bibnamefont{S{\o}rensen}},
  \bibinfo{author}{\bibfnamefont{K.}~\bibnamefont{Usami}},
  \bibinfo{author}{\bibfnamefont{A.}~\bibnamefont{Schliesser}},
  \bibnamefont{et~al.}, \bibinfo{journal}{Nature}
  \textbf{\bibinfo{volume}{507}}, \bibinfo{pages}{81} (\bibinfo{year}{2014}),
  ISSN \bibinfo{issn}{1476-4687}.

\bibitem[{\citenamefont{Andrews et~al.}(2014)\citenamefont{Andrews, Peterson,
  Purdy, Cicak, Simmonds, Regal, and
  Lehnert}}]{andrewsBidirectionalEfficientConversion2014a}
\bibinfo{author}{\bibfnamefont{R.~W.} \bibnamefont{Andrews}},
  \bibinfo{author}{\bibfnamefont{R.~W.} \bibnamefont{Peterson}},
  \bibinfo{author}{\bibfnamefont{T.~P.} \bibnamefont{Purdy}},
  \bibinfo{author}{\bibfnamefont{K.}~\bibnamefont{Cicak}},
  \bibinfo{author}{\bibfnamefont{R.~W.} \bibnamefont{Simmonds}},
  \bibinfo{author}{\bibfnamefont{C.~A.} \bibnamefont{Regal}}, \bibnamefont{and}
  \bibinfo{author}{\bibfnamefont{K.~W.} \bibnamefont{Lehnert}},
  \bibinfo{journal}{Nature Physics} \textbf{\bibinfo{volume}{10}},
  \bibinfo{pages}{321} (\bibinfo{year}{2014}), ISSN \bibinfo{issn}{1745-2481}.

\bibitem[{\citenamefont{Wang and Clerk}(2012)}]{wangUsingInterferenceHigh2012a}
\bibinfo{author}{\bibfnamefont{Y.-D.} \bibnamefont{Wang}} \bibnamefont{and}
  \bibinfo{author}{\bibfnamefont{A.~A.} \bibnamefont{Clerk}},
  \bibinfo{journal}{Physical Review Letters} \textbf{\bibinfo{volume}{108}},
  \bibinfo{pages}{153603} (\bibinfo{year}{2012}).

\bibitem[{\citenamefont{Tian}(2012)}]{tianAdiabaticStateConversion2012a}
\bibinfo{author}{\bibfnamefont{L.}~\bibnamefont{Tian}},
  \bibinfo{journal}{Physical Review Letters} \textbf{\bibinfo{volume}{108}},
  \bibinfo{pages}{153604} (\bibinfo{year}{2012}).

\bibitem[{\citenamefont{Hill et~al.}(2012)\citenamefont{Hill, {Safavi-Naeini},
  Chan, and Painter}}]{hillCoherentOpticalWavelength2012a}
\bibinfo{author}{\bibfnamefont{J.~T.} \bibnamefont{Hill}},
  \bibinfo{author}{\bibfnamefont{A.~H.} \bibnamefont{{Safavi-Naeini}}},
  \bibinfo{author}{\bibfnamefont{J.}~\bibnamefont{Chan}}, \bibnamefont{and}
  \bibinfo{author}{\bibfnamefont{O.}~\bibnamefont{Painter}},
  \bibinfo{journal}{Nature Communications} \textbf{\bibinfo{volume}{3}},
  \bibinfo{pages}{1196} (\bibinfo{year}{2012}), ISSN \bibinfo{issn}{2041-1723}.

\bibitem[{\citenamefont{Barzanjeh et~al.}(2012)\citenamefont{Barzanjeh, Abdi,
  Milburn, Tombesi, and
  Vitali}}]{barzanjehReversibleOpticaltoMicrowaveQuantum2012a}
\bibinfo{author}{\bibfnamefont{S.}~\bibnamefont{Barzanjeh}},
  \bibinfo{author}{\bibfnamefont{M.}~\bibnamefont{Abdi}},
  \bibinfo{author}{\bibfnamefont{G.~J.} \bibnamefont{Milburn}},
  \bibinfo{author}{\bibfnamefont{P.}~\bibnamefont{Tombesi}}, \bibnamefont{and}
  \bibinfo{author}{\bibfnamefont{D.}~\bibnamefont{Vitali}},
  \bibinfo{journal}{Physical Review Letters} \textbf{\bibinfo{volume}{109}},
  \bibinfo{pages}{130503} (\bibinfo{year}{2012}).

\bibitem[{\citenamefont{{Safavi-Naeini} and
  Painter}(2011)}]{safavi-naeiniProposalOptomechanicalTraveling2011a}
\bibinfo{author}{\bibfnamefont{A.~H.} \bibnamefont{{Safavi-Naeini}}}
  \bibnamefont{and} \bibinfo{author}{\bibfnamefont{O.}~\bibnamefont{Painter}},
  \bibinfo{journal}{New Journal of Physics} \textbf{\bibinfo{volume}{13}},
  \bibinfo{pages}{013017} (\bibinfo{year}{2011}), ISSN
  \bibinfo{issn}{1367-2630}.

\bibitem[{\citenamefont{Tian and
  Wang}(2010)}]{tianOpticalWavelengthConversion2010a}
\bibinfo{author}{\bibfnamefont{L.}~\bibnamefont{Tian}} \bibnamefont{and}
  \bibinfo{author}{\bibfnamefont{H.}~\bibnamefont{Wang}},
  \bibinfo{journal}{Physical Review A} \textbf{\bibinfo{volume}{82}},
  \bibinfo{pages}{053806} (\bibinfo{year}{2010}).

\bibitem[{\citenamefont{Stannigel et~al.}(2010)\citenamefont{Stannigel, Rabl,
  S{\o}rensen, Zoller, and
  Lukin}}]{stannigelOptomechanicalTransducersLongDistance2010a}
\bibinfo{author}{\bibfnamefont{K.}~\bibnamefont{Stannigel}},
  \bibinfo{author}{\bibfnamefont{P.}~\bibnamefont{Rabl}},
  \bibinfo{author}{\bibfnamefont{A.~S.} \bibnamefont{S{\o}rensen}},
  \bibinfo{author}{\bibfnamefont{P.}~\bibnamefont{Zoller}}, \bibnamefont{and}
  \bibinfo{author}{\bibfnamefont{M.~D.} \bibnamefont{Lukin}},
  \bibinfo{journal}{Physical Review Letters} \textbf{\bibinfo{volume}{105}},
  \bibinfo{pages}{220501} (\bibinfo{year}{2010}).

\bibitem[{\citenamefont{Bartholomew et~al.}(2020)\citenamefont{Bartholomew,
  Rochman, Xie, Kindem, Ruskuc, Craiciu, Lei, and
  Faraon}}]{bartholomewOnchipCoherentMicrowavetooptical2020a}
\bibinfo{author}{\bibfnamefont{J.~G.} \bibnamefont{Bartholomew}},
  \bibinfo{author}{\bibfnamefont{J.}~\bibnamefont{Rochman}},
  \bibinfo{author}{\bibfnamefont{T.}~\bibnamefont{Xie}},
  \bibinfo{author}{\bibfnamefont{J.~M.} \bibnamefont{Kindem}},
  \bibinfo{author}{\bibfnamefont{A.}~\bibnamefont{Ruskuc}},
  \bibinfo{author}{\bibfnamefont{I.}~\bibnamefont{Craiciu}},
  \bibinfo{author}{\bibfnamefont{M.}~\bibnamefont{Lei}}, \bibnamefont{and}
  \bibinfo{author}{\bibfnamefont{A.}~\bibnamefont{Faraon}},
  \bibinfo{journal}{Nature Communications} \textbf{\bibinfo{volume}{11}},
  \bibinfo{pages}{3266} (\bibinfo{year}{2020}), ISSN \bibinfo{issn}{2041-1723},
  \eprint{1912.03671}.

\bibitem[{\citenamefont{Barnett and
  Longdell}(2020)}]{barnettTheoryMicrowaveOpticalConversion2020}
\bibinfo{author}{\bibfnamefont{P.~S.} \bibnamefont{Barnett}} \bibnamefont{and}
  \bibinfo{author}{\bibfnamefont{J.~J.} \bibnamefont{Longdell}},
  \bibinfo{journal}{arXiv:2008.10834 [quant-ph]}  (\bibinfo{year}{2020}),
  \bibinfo{note}{comment: 10 pages, 7 figures}, \eprint{2008.10834}.

\bibitem[{\citenamefont{Welinski et~al.}(2019)\citenamefont{Welinski, Woodburn,
  Lauk, Cone, Simon, Goldner, and Thiel}}]{welinskiElectronSpinCoherence2019}
\bibinfo{author}{\bibfnamefont{S.}~\bibnamefont{Welinski}},
  \bibinfo{author}{\bibfnamefont{P.~J.~T.} \bibnamefont{Woodburn}},
  \bibinfo{author}{\bibfnamefont{N.}~\bibnamefont{Lauk}},
  \bibinfo{author}{\bibfnamefont{R.~L.} \bibnamefont{Cone}},
  \bibinfo{author}{\bibfnamefont{C.}~\bibnamefont{Simon}},
  \bibinfo{author}{\bibfnamefont{P.}~\bibnamefont{Goldner}}, \bibnamefont{and}
  \bibinfo{author}{\bibfnamefont{C.~W.} \bibnamefont{Thiel}},
  \bibinfo{journal}{Physical Review Letters} \textbf{\bibinfo{volume}{122}},
  \bibinfo{pages}{247401} (\bibinfo{year}{2019}).

\bibitem[{\citenamefont{Vogt et~al.}(2019)\citenamefont{Vogt, Gross, Han, Pal,
  Lam, Kiffner, and Li}}]{vogtEfficientMicrowavetoopticalConversion2019a}
\bibinfo{author}{\bibfnamefont{T.}~\bibnamefont{Vogt}},
  \bibinfo{author}{\bibfnamefont{C.}~\bibnamefont{Gross}},
  \bibinfo{author}{\bibfnamefont{J.}~\bibnamefont{Han}},
  \bibinfo{author}{\bibfnamefont{S.~B.} \bibnamefont{Pal}},
  \bibinfo{author}{\bibfnamefont{M.}~\bibnamefont{Lam}},
  \bibinfo{author}{\bibfnamefont{M.}~\bibnamefont{Kiffner}}, \bibnamefont{and}
  \bibinfo{author}{\bibfnamefont{W.}~\bibnamefont{Li}},
  \bibinfo{journal}{Physical Review A} \textbf{\bibinfo{volume}{99}},
  \bibinfo{pages}{023832} (\bibinfo{year}{2019}).

\bibitem[{\citenamefont{Petrosyan et~al.}(2019)\citenamefont{Petrosyan,
  M{\o}lmer, Fort{\'a}gh, and
  Saffman}}]{petrosyanMicrowaveOpticalConversion2019}
\bibinfo{author}{\bibfnamefont{D.}~\bibnamefont{Petrosyan}},
  \bibinfo{author}{\bibfnamefont{K.}~\bibnamefont{M{\o}lmer}},
  \bibinfo{author}{\bibfnamefont{J.}~\bibnamefont{Fort{\'a}gh}},
  \bibnamefont{and} \bibinfo{author}{\bibfnamefont{M.}~\bibnamefont{Saffman}},
  \bibinfo{journal}{New Journal of Physics} \textbf{\bibinfo{volume}{21}},
  \bibinfo{pages}{073033} (\bibinfo{year}{2019}), ISSN
  \bibinfo{issn}{1367-2630}.

\bibitem[{\citenamefont{{Fernandez-Gonzalvo}
  et~al.}(2019)\citenamefont{{Fernandez-Gonzalvo}, Horvath, Chen, and
  Longdell}}]{fernandez-gonzalvoCavityenhancedRamanHeterodyne2019a}
\bibinfo{author}{\bibfnamefont{X.}~\bibnamefont{{Fernandez-Gonzalvo}}},
  \bibinfo{author}{\bibfnamefont{S.~P.} \bibnamefont{Horvath}},
  \bibinfo{author}{\bibfnamefont{Y.-H.} \bibnamefont{Chen}}, \bibnamefont{and}
  \bibinfo{author}{\bibfnamefont{J.~J.} \bibnamefont{Longdell}},
  \bibinfo{journal}{Physical Review A} \textbf{\bibinfo{volume}{100}},
  \bibinfo{pages}{033807} (\bibinfo{year}{2019}).

\bibitem[{\citenamefont{Covey et~al.}(2019)\citenamefont{Covey, Sipahigil, and
  Saffman}}]{coveyMicrowavetoopticalConversionFourwave2019}
\bibinfo{author}{\bibfnamefont{J.~P.} \bibnamefont{Covey}},
  \bibinfo{author}{\bibfnamefont{A.}~\bibnamefont{Sipahigil}},
  \bibnamefont{and} \bibinfo{author}{\bibfnamefont{M.}~\bibnamefont{Saffman}},
  \bibinfo{journal}{Physical Review A} \textbf{\bibinfo{volume}{100}},
  \bibinfo{pages}{012307} (\bibinfo{year}{2019}).

\bibitem[{\citenamefont{Han et~al.}(2018)\citenamefont{Han, Vogt, Gross,
  Jaksch, Kiffner, and Li}}]{hanCoherentMicrowavetoOpticalConversion2018a}
\bibinfo{author}{\bibfnamefont{J.}~\bibnamefont{Han}},
  \bibinfo{author}{\bibfnamefont{T.}~\bibnamefont{Vogt}},
  \bibinfo{author}{\bibfnamefont{C.}~\bibnamefont{Gross}},
  \bibinfo{author}{\bibfnamefont{D.}~\bibnamefont{Jaksch}},
  \bibinfo{author}{\bibfnamefont{M.}~\bibnamefont{Kiffner}}, \bibnamefont{and}
  \bibinfo{author}{\bibfnamefont{W.}~\bibnamefont{Li}},
  \bibinfo{journal}{Physical Review Letters} \textbf{\bibinfo{volume}{120}},
  \bibinfo{pages}{093201} (\bibinfo{year}{2018}), ISSN
  \bibinfo{issn}{0031-9007, 1079-7114}.

\bibitem[{\citenamefont{Gard et~al.}(2017)\citenamefont{Gard, Jacobs,
  McDermott, and Saffman}}]{gardMicrowavetoopticalFrequencyConversion2017a}
\bibinfo{author}{\bibfnamefont{B.~T.} \bibnamefont{Gard}},
  \bibinfo{author}{\bibfnamefont{K.}~\bibnamefont{Jacobs}},
  \bibinfo{author}{\bibfnamefont{R.}~\bibnamefont{McDermott}},
  \bibnamefont{and} \bibinfo{author}{\bibfnamefont{M.}~\bibnamefont{Saffman}},
  \bibinfo{journal}{Physical Review A} \textbf{\bibinfo{volume}{96}},
  \bibinfo{pages}{013833} (\bibinfo{year}{2017}), ISSN
  \bibinfo{issn}{2469-9926, 2469-9934}.

\bibitem[{\citenamefont{Kiffner et~al.}(2016)\citenamefont{Kiffner, Feizpour,
  Kaczmarek, Jaksch, and
  Nunn}}]{kiffnerTwowayInterconversionMillimeterwave2016a}
\bibinfo{author}{\bibfnamefont{M.}~\bibnamefont{Kiffner}},
  \bibinfo{author}{\bibfnamefont{A.}~\bibnamefont{Feizpour}},
  \bibinfo{author}{\bibfnamefont{K.~T.} \bibnamefont{Kaczmarek}},
  \bibinfo{author}{\bibfnamefont{D.}~\bibnamefont{Jaksch}}, \bibnamefont{and}
  \bibinfo{author}{\bibfnamefont{J.}~\bibnamefont{Nunn}}, \bibinfo{journal}{New
  Journal of Physics} \textbf{\bibinfo{volume}{18}}, \bibinfo{pages}{093030}
  (\bibinfo{year}{2016}), ISSN \bibinfo{issn}{1367-2630}.

\bibitem[{\citenamefont{{Fernandez-Gonzalvo}
  et~al.}(2015)\citenamefont{{Fernandez-Gonzalvo}, Chen, Yin, Rogge, and
  Longdell}}]{fernandez-gonzalvoCoherentFrequencyUpconversion2015}
\bibinfo{author}{\bibfnamefont{X.}~\bibnamefont{{Fernandez-Gonzalvo}}},
  \bibinfo{author}{\bibfnamefont{Y.-H.} \bibnamefont{Chen}},
  \bibinfo{author}{\bibfnamefont{C.}~\bibnamefont{Yin}},
  \bibinfo{author}{\bibfnamefont{S.}~\bibnamefont{Rogge}}, \bibnamefont{and}
  \bibinfo{author}{\bibfnamefont{J.~J.} \bibnamefont{Longdell}},
  \bibinfo{journal}{Physical Review A} \textbf{\bibinfo{volume}{92}},
  \bibinfo{pages}{062313} (\bibinfo{year}{2015}).

\bibitem[{\citenamefont{Williamson et~al.}(2014)\citenamefont{Williamson, Chen,
  and Longdell}}]{williamsonMagnetoOpticModulatorUnit2014a}
\bibinfo{author}{\bibfnamefont{L.~A.} \bibnamefont{Williamson}},
  \bibinfo{author}{\bibfnamefont{Y.-H.} \bibnamefont{Chen}}, \bibnamefont{and}
  \bibinfo{author}{\bibfnamefont{J.~J.} \bibnamefont{Longdell}},
  \bibinfo{journal}{Physical Review Letters} \textbf{\bibinfo{volume}{113}},
  \bibinfo{pages}{203601} (\bibinfo{year}{2014}).

\bibitem[{\citenamefont{O'Brien et~al.}(2014)\citenamefont{O'Brien, Lauk, Blum,
  Morigi, and Fleischhauer}}]{obrienInterfacingSuperconductingQubits2014}
\bibinfo{author}{\bibfnamefont{C.}~\bibnamefont{O'Brien}},
  \bibinfo{author}{\bibfnamefont{N.}~\bibnamefont{Lauk}},
  \bibinfo{author}{\bibfnamefont{S.}~\bibnamefont{Blum}},
  \bibinfo{author}{\bibfnamefont{G.}~\bibnamefont{Morigi}}, \bibnamefont{and}
  \bibinfo{author}{\bibfnamefont{M.}~\bibnamefont{Fleischhauer}},
  \bibinfo{journal}{Physical Review Letters} \textbf{\bibinfo{volume}{113}},
  \bibinfo{pages}{063603} (\bibinfo{year}{2014}), ISSN
  \bibinfo{issn}{0031-9007, 1079-7114}.

\bibitem[{\citenamefont{Hafezi et~al.}(2012)\citenamefont{Hafezi, Kim, Rolston,
  Orozco, Lev, and Taylor}}]{hafeziAtomicInterfaceMicrowave2012}
\bibinfo{author}{\bibfnamefont{M.}~\bibnamefont{Hafezi}},
  \bibinfo{author}{\bibfnamefont{Z.}~\bibnamefont{Kim}},
  \bibinfo{author}{\bibfnamefont{S.~L.} \bibnamefont{Rolston}},
  \bibinfo{author}{\bibfnamefont{L.~A.} \bibnamefont{Orozco}},
  \bibinfo{author}{\bibfnamefont{B.~L.} \bibnamefont{Lev}}, \bibnamefont{and}
  \bibinfo{author}{\bibfnamefont{J.~M.} \bibnamefont{Taylor}},
  \bibinfo{journal}{Physical Review A} \textbf{\bibinfo{volume}{85}},
  \bibinfo{pages}{020302} (\bibinfo{year}{2012}), ISSN
  \bibinfo{issn}{1050-2947, 1094-1622}.

\bibitem[{\citenamefont{Verd{\'u} et~al.}(2009)\citenamefont{Verd{\'u}, Zoubi,
  Koller, Majer, Ritsch, and Schmiedmayer}}]{verduStrongMagneticCoupling2009a}
\bibinfo{author}{\bibfnamefont{J.}~\bibnamefont{Verd{\'u}}},
  \bibinfo{author}{\bibfnamefont{H.}~\bibnamefont{Zoubi}},
  \bibinfo{author}{\bibfnamefont{C.}~\bibnamefont{Koller}},
  \bibinfo{author}{\bibfnamefont{J.}~\bibnamefont{Majer}},
  \bibinfo{author}{\bibfnamefont{H.}~\bibnamefont{Ritsch}}, \bibnamefont{and}
  \bibinfo{author}{\bibfnamefont{J.}~\bibnamefont{Schmiedmayer}},
  \bibinfo{journal}{Physical Review Letters} \textbf{\bibinfo{volume}{103}},
  \bibinfo{pages}{043603} (\bibinfo{year}{2009}).

\bibitem[{\citenamefont{Imamo{\u g}lu}(2009)}]{imamogluCavityQEDBased2009a}
\bibinfo{author}{\bibfnamefont{A.}~\bibnamefont{Imamo{\u g}lu}},
  \bibinfo{journal}{Physical Review Letters} \textbf{\bibinfo{volume}{102}},
  \bibinfo{pages}{083602} (\bibinfo{year}{2009}).

\bibitem[{\citenamefont{Zhu et~al.}(2011)\citenamefont{Zhu, Saito, Kemp,
  Kakuyanagi, Karimoto, Nakano, Munro, Tokura, Everitt, Nemoto
  et~al.}}]{zhuCoherentCouplingSuperconducting2011}
\bibinfo{author}{\bibfnamefont{X.}~\bibnamefont{Zhu}},
  \bibinfo{author}{\bibfnamefont{S.}~\bibnamefont{Saito}},
  \bibinfo{author}{\bibfnamefont{A.}~\bibnamefont{Kemp}},
  \bibinfo{author}{\bibfnamefont{K.}~\bibnamefont{Kakuyanagi}},
  \bibinfo{author}{\bibfnamefont{S.-i.} \bibnamefont{Karimoto}},
  \bibinfo{author}{\bibfnamefont{H.}~\bibnamefont{Nakano}},
  \bibinfo{author}{\bibfnamefont{W.~J.} \bibnamefont{Munro}},
  \bibinfo{author}{\bibfnamefont{Y.}~\bibnamefont{Tokura}},
  \bibinfo{author}{\bibfnamefont{M.~S.} \bibnamefont{Everitt}},
  \bibinfo{author}{\bibfnamefont{K.}~\bibnamefont{Nemoto}},
  \bibnamefont{et~al.}, \bibinfo{journal}{Nature}
  \textbf{\bibinfo{volume}{478}}, \bibinfo{pages}{221} (\bibinfo{year}{2011}),
  ISSN \bibinfo{issn}{0028-0836, 1476-4687}.

\bibitem[{\citenamefont{Dold et~al.}(2019)\citenamefont{Dold, Zollitsch,
  O'Sullivan, Welinski, Ferrier, Goldner, {de Graaf}, Lindstr{\"o}m, and
  Morton}}]{doldHighCooperativityCouplingRareEarth2019}
\bibinfo{author}{\bibfnamefont{G.}~\bibnamefont{Dold}},
  \bibinfo{author}{\bibfnamefont{C.~W.} \bibnamefont{Zollitsch}},
  \bibinfo{author}{\bibfnamefont{J.}~\bibnamefont{O'Sullivan}},
  \bibinfo{author}{\bibfnamefont{S.}~\bibnamefont{Welinski}},
  \bibinfo{author}{\bibfnamefont{A.}~\bibnamefont{Ferrier}},
  \bibinfo{author}{\bibfnamefont{P.}~\bibnamefont{Goldner}},
  \bibinfo{author}{\bibfnamefont{S.}~\bibnamefont{{de Graaf}}},
  \bibinfo{author}{\bibfnamefont{T.}~\bibnamefont{Lindstr{\"o}m}},
  \bibnamefont{and} \bibinfo{author}{\bibfnamefont{J.~J.}
  \bibnamefont{Morton}}, \bibinfo{journal}{Physical Review Applied}
  \textbf{\bibinfo{volume}{11}}, \bibinfo{pages}{054082}
  (\bibinfo{year}{2019}).

\bibitem[{\citenamefont{Dory et~al.}(2019)\citenamefont{Dory, Vercruysse, Yang,
  Sapra, Rugar, Sun, Lukin, Piggott, Zhang, Radulaski
  et~al.}}]{doryInversedesignedDiamondPhotonics2019}
\bibinfo{author}{\bibfnamefont{C.}~\bibnamefont{Dory}},
  \bibinfo{author}{\bibfnamefont{D.}~\bibnamefont{Vercruysse}},
  \bibinfo{author}{\bibfnamefont{K.~Y.} \bibnamefont{Yang}},
  \bibinfo{author}{\bibfnamefont{N.~V.} \bibnamefont{Sapra}},
  \bibinfo{author}{\bibfnamefont{A.~E.} \bibnamefont{Rugar}},
  \bibinfo{author}{\bibfnamefont{S.}~\bibnamefont{Sun}},
  \bibinfo{author}{\bibfnamefont{D.~M.} \bibnamefont{Lukin}},
  \bibinfo{author}{\bibfnamefont{A.~Y.} \bibnamefont{Piggott}},
  \bibinfo{author}{\bibfnamefont{J.~L.} \bibnamefont{Zhang}},
  \bibinfo{author}{\bibfnamefont{M.}~\bibnamefont{Radulaski}},
  \bibnamefont{et~al.}, \bibinfo{journal}{Nature Communications}
  \textbf{\bibinfo{volume}{10}}, \bibinfo{pages}{3309} (\bibinfo{year}{2019}),
  ISSN \bibinfo{issn}{2041-1723}.

\bibitem[{\citenamefont{Lukin et~al.}(2020{\natexlab{a}})\citenamefont{Lukin,
  Dory, Guidry, Yang, Mishra, Trivedi, Radulaski, Sun, Vercruysse, Ahn
  et~al.}}]{lukin4HsiliconcarbideoninsulatorIntegratedQuantum2020}
\bibinfo{author}{\bibfnamefont{D.~M.} \bibnamefont{Lukin}},
  \bibinfo{author}{\bibfnamefont{C.}~\bibnamefont{Dory}},
  \bibinfo{author}{\bibfnamefont{M.~A.} \bibnamefont{Guidry}},
  \bibinfo{author}{\bibfnamefont{K.~Y.} \bibnamefont{Yang}},
  \bibinfo{author}{\bibfnamefont{S.~D.} \bibnamefont{Mishra}},
  \bibinfo{author}{\bibfnamefont{R.}~\bibnamefont{Trivedi}},
  \bibinfo{author}{\bibfnamefont{M.}~\bibnamefont{Radulaski}},
  \bibinfo{author}{\bibfnamefont{S.}~\bibnamefont{Sun}},
  \bibinfo{author}{\bibfnamefont{D.}~\bibnamefont{Vercruysse}},
  \bibinfo{author}{\bibfnamefont{G.~H.} \bibnamefont{Ahn}},
  \bibnamefont{et~al.}, \bibinfo{journal}{Nature Photonics}
  \textbf{\bibinfo{volume}{14}}, \bibinfo{pages}{330}
  (\bibinfo{year}{2020}{\natexlab{a}}), ISSN \bibinfo{issn}{1749-4893}.

\bibitem[{\citenamefont{Wan et~al.}(2020)\citenamefont{Wan, Lu, Chen, Walsh,
  Trusheim, De~Santis, Bersin, Harris, Mouradian, Christen
  et~al.}}]{wanLargescaleIntegrationArtificial2020}
\bibinfo{author}{\bibfnamefont{N.~H.} \bibnamefont{Wan}},
  \bibinfo{author}{\bibfnamefont{T.-J.} \bibnamefont{Lu}},
  \bibinfo{author}{\bibfnamefont{K.~C.} \bibnamefont{Chen}},
  \bibinfo{author}{\bibfnamefont{M.~P.} \bibnamefont{Walsh}},
  \bibinfo{author}{\bibfnamefont{M.~E.} \bibnamefont{Trusheim}},
  \bibinfo{author}{\bibfnamefont{L.}~\bibnamefont{De~Santis}},
  \bibinfo{author}{\bibfnamefont{E.~A.} \bibnamefont{Bersin}},
  \bibinfo{author}{\bibfnamefont{I.~B.} \bibnamefont{Harris}},
  \bibinfo{author}{\bibfnamefont{S.~L.} \bibnamefont{Mouradian}},
  \bibinfo{author}{\bibfnamefont{I.~R.} \bibnamefont{Christen}},
  \bibnamefont{et~al.}, \bibinfo{journal}{Nature}
  \textbf{\bibinfo{volume}{583}}, \bibinfo{pages}{226} (\bibinfo{year}{2020}),
  ISSN \bibinfo{issn}{1476-4687}.

\bibitem[{\citenamefont{Dicke}(1954)}]{dickeCoherenceSpontaneousRadiation1954}
\bibinfo{author}{\bibfnamefont{R.~H.} \bibnamefont{Dicke}},
  \bibinfo{journal}{Physical Review} \textbf{\bibinfo{volume}{93}},
  \bibinfo{pages}{99} (\bibinfo{year}{1954}).

\bibitem[{\citenamefont{Gross and
  Haroche}(1982)}]{grossSuperradianceEssayTheory1982}
\bibinfo{author}{\bibfnamefont{M.}~\bibnamefont{Gross}} \bibnamefont{and}
  \bibinfo{author}{\bibfnamefont{S.}~\bibnamefont{Haroche}},
  \bibinfo{journal}{Physics Reports} \textbf{\bibinfo{volume}{93}},
  \bibinfo{pages}{301} (\bibinfo{year}{1982}), ISSN \bibinfo{issn}{0370-1573}.

\bibitem[{\citenamefont{Trivedi et~al.}(2019)\citenamefont{Trivedi, Radulaski,
  Fischer, Fan, and Vu{\v c}kovi{\'c}}}]{trivediPhotonBlockadeWeakly2019}
\bibinfo{author}{\bibfnamefont{R.}~\bibnamefont{Trivedi}},
  \bibinfo{author}{\bibfnamefont{M.}~\bibnamefont{Radulaski}},
  \bibinfo{author}{\bibfnamefont{K.~A.} \bibnamefont{Fischer}},
  \bibinfo{author}{\bibfnamefont{S.}~\bibnamefont{Fan}}, \bibnamefont{and}
  \bibinfo{author}{\bibfnamefont{J.}~\bibnamefont{Vu{\v c}kovi{\'c}}},
  \bibinfo{journal}{Physical Review Letters} \textbf{\bibinfo{volume}{122}},
  \bibinfo{pages}{243602} (\bibinfo{year}{2019}).

\bibitem[{\citenamefont{Duan et~al.}(2001)\citenamefont{Duan, Lukin, Cirac, and
  Zoller}}]{duanLongdistanceQuantumCommunication2001a}
\bibinfo{author}{\bibfnamefont{L.-M.} \bibnamefont{Duan}},
  \bibinfo{author}{\bibfnamefont{M.~D.} \bibnamefont{Lukin}},
  \bibinfo{author}{\bibfnamefont{J.~I.} \bibnamefont{Cirac}}, \bibnamefont{and}
  \bibinfo{author}{\bibfnamefont{P.}~\bibnamefont{Zoller}},
  \bibinfo{journal}{Nature} \textbf{\bibinfo{volume}{414}},
  \bibinfo{pages}{413} (\bibinfo{year}{2001}), ISSN \bibinfo{issn}{1476-4687}.

\bibitem[{\citenamefont{{Gonz{\'a}lez-Tudela}
  et~al.}(2015)\citenamefont{{Gonz{\'a}lez-Tudela}, Paulisch, Chang, Kimble,
  and Cirac}}]{gonzalez-tudelaDeterministicGenerationArbitrary2015}
\bibinfo{author}{\bibfnamefont{A.}~\bibnamefont{{Gonz{\'a}lez-Tudela}}},
  \bibinfo{author}{\bibfnamefont{V.}~\bibnamefont{Paulisch}},
  \bibinfo{author}{\bibfnamefont{D.~E.} \bibnamefont{Chang}},
  \bibinfo{author}{\bibfnamefont{H.~J.} \bibnamefont{Kimble}},
  \bibnamefont{and} \bibinfo{author}{\bibfnamefont{J.~I.} \bibnamefont{Cirac}},
  \bibinfo{journal}{Physical Review Letters} \textbf{\bibinfo{volume}{115}},
  \bibinfo{pages}{163603} (\bibinfo{year}{2015}).

\bibitem[{\citenamefont{Paulisch et~al.}(2019)\citenamefont{Paulisch,
  {Perarnau-Llobet}, {Gonz{\'a}lez-Tudela}, and
  Cirac}}]{paulischQuantumMetrologyOnedimensional2019}
\bibinfo{author}{\bibfnamefont{V.}~\bibnamefont{Paulisch}},
  \bibinfo{author}{\bibfnamefont{M.}~\bibnamefont{{Perarnau-Llobet}}},
  \bibinfo{author}{\bibfnamefont{A.}~\bibnamefont{{Gonz{\'a}lez-Tudela}}},
  \bibnamefont{and} \bibinfo{author}{\bibfnamefont{J.~I.} \bibnamefont{Cirac}},
  \bibinfo{journal}{Physical Review A} \textbf{\bibinfo{volume}{99}},
  \bibinfo{pages}{043807} (\bibinfo{year}{2019}).

\bibitem[{\citenamefont{Evans et~al.}(2016)\citenamefont{Evans, Sipahigil,
  Sukachev, Zibrov, and Lukin}}]{evansNarrowLinewidthHomogeneousOptical2016}
\bibinfo{author}{\bibfnamefont{R.~E.} \bibnamefont{Evans}},
  \bibinfo{author}{\bibfnamefont{A.}~\bibnamefont{Sipahigil}},
  \bibinfo{author}{\bibfnamefont{D.~D.} \bibnamefont{Sukachev}},
  \bibinfo{author}{\bibfnamefont{A.~S.} \bibnamefont{Zibrov}},
  \bibnamefont{and} \bibinfo{author}{\bibfnamefont{M.~D.} \bibnamefont{Lukin}},
  \bibinfo{journal}{Physical Review Applied} \textbf{\bibinfo{volume}{5}},
  \bibinfo{pages}{044010} (\bibinfo{year}{2016}), ISSN
  \bibinfo{issn}{2331-7019}.

\bibitem[{\citenamefont{Dibos et~al.}(2018)\citenamefont{Dibos, Raha, Phenicie,
  and Thompson}}]{dibosAtomicSourceSingle2018a}
\bibinfo{author}{\bibfnamefont{A.~M.} \bibnamefont{Dibos}},
  \bibinfo{author}{\bibfnamefont{M.}~\bibnamefont{Raha}},
  \bibinfo{author}{\bibfnamefont{C.~M.} \bibnamefont{Phenicie}},
  \bibnamefont{and} \bibinfo{author}{\bibfnamefont{J.~D.}
  \bibnamefont{Thompson}}, \bibinfo{journal}{Physical Review Letters}
  \textbf{\bibinfo{volume}{120}}, \bibinfo{pages}{243601}
  (\bibinfo{year}{2018}), ISSN \bibinfo{issn}{0031-9007, 1079-7114}.

\bibitem[{\citenamefont{Zhong et~al.}(2018)\citenamefont{Zhong, Kindem,
  Bartholomew, Rochman, Craiciu, Verma, Nam, Marsili, Shaw, Beyer
  et~al.}}]{zhongOpticallyAddressingSingle2018}
\bibinfo{author}{\bibfnamefont{T.}~\bibnamefont{Zhong}},
  \bibinfo{author}{\bibfnamefont{J.~M.} \bibnamefont{Kindem}},
  \bibinfo{author}{\bibfnamefont{J.~G.} \bibnamefont{Bartholomew}},
  \bibinfo{author}{\bibfnamefont{J.}~\bibnamefont{Rochman}},
  \bibinfo{author}{\bibfnamefont{I.}~\bibnamefont{Craiciu}},
  \bibinfo{author}{\bibfnamefont{V.}~\bibnamefont{Verma}},
  \bibinfo{author}{\bibfnamefont{S.~W.} \bibnamefont{Nam}},
  \bibinfo{author}{\bibfnamefont{F.}~\bibnamefont{Marsili}},
  \bibinfo{author}{\bibfnamefont{M.~D.} \bibnamefont{Shaw}},
  \bibinfo{author}{\bibfnamefont{A.~D.} \bibnamefont{Beyer}},
  \bibnamefont{et~al.}, \bibinfo{journal}{Physical Review Letters}
  \textbf{\bibinfo{volume}{121}}, \bibinfo{pages}{183603}
  (\bibinfo{year}{2018}), ISSN \bibinfo{issn}{0031-9007, 1079-7114}.

\bibitem[{\citenamefont{Dong and
  Petersen}(2010)}]{dongQuantumControlTheory2010}
\bibinfo{author}{\bibfnamefont{D.}~\bibnamefont{Dong}} \bibnamefont{and}
  \bibinfo{author}{\bibfnamefont{I.~R.} \bibnamefont{Petersen}},
  \bibinfo{journal}{IET Control Theory Applications}
  \textbf{\bibinfo{volume}{4}}, \bibinfo{pages}{2651} (\bibinfo{year}{2010}),
  ISSN \bibinfo{issn}{1751-8652}.

\bibitem[{\citenamefont{Koch}(2016)}]{kochControllingOpenQuantum2016}
\bibinfo{author}{\bibfnamefont{C.~P.} \bibnamefont{Koch}},
  \bibinfo{journal}{Journal of Physics: Condensed Matter}
  \textbf{\bibinfo{volume}{28}}, \bibinfo{pages}{213001}
  (\bibinfo{year}{2016}), ISSN \bibinfo{issn}{0953-8984, 1361-648X}.

\bibitem[{\citenamefont{Grzesiak et~al.}(2020)\citenamefont{Grzesiak,
  Bl{\"u}mel, Wright, Beck, Pisenti, Li, Chaplin, Amini, Debnath, Chen
  et~al.}}]{grzesiakEfficientArbitrarySimultaneously2020}
\bibinfo{author}{\bibfnamefont{N.}~\bibnamefont{Grzesiak}},
  \bibinfo{author}{\bibfnamefont{R.}~\bibnamefont{Bl{\"u}mel}},
  \bibinfo{author}{\bibfnamefont{K.}~\bibnamefont{Wright}},
  \bibinfo{author}{\bibfnamefont{K.~M.} \bibnamefont{Beck}},
  \bibinfo{author}{\bibfnamefont{N.~C.} \bibnamefont{Pisenti}},
  \bibinfo{author}{\bibfnamefont{M.}~\bibnamefont{Li}},
  \bibinfo{author}{\bibfnamefont{V.}~\bibnamefont{Chaplin}},
  \bibinfo{author}{\bibfnamefont{J.~M.} \bibnamefont{Amini}},
  \bibinfo{author}{\bibfnamefont{S.}~\bibnamefont{Debnath}},
  \bibinfo{author}{\bibfnamefont{J.-S.} \bibnamefont{Chen}},
  \bibnamefont{et~al.}, \bibinfo{journal}{Nature Communications}
  \textbf{\bibinfo{volume}{11}}, \bibinfo{pages}{2963} (\bibinfo{year}{2020}),
  ISSN \bibinfo{issn}{2041-1723}.

\bibitem[{\citenamefont{Poulsen et~al.}(2010)\citenamefont{Poulsen, Sklarz,
  Tannor, and Calarco}}]{poulsenCorrectingErrorsQuantum2010}
\bibinfo{author}{\bibfnamefont{U.~V.} \bibnamefont{Poulsen}},
  \bibinfo{author}{\bibfnamefont{S.}~\bibnamefont{Sklarz}},
  \bibinfo{author}{\bibfnamefont{D.}~\bibnamefont{Tannor}}, \bibnamefont{and}
  \bibinfo{author}{\bibfnamefont{T.}~\bibnamefont{Calarco}},
  \bibinfo{journal}{Physical Review A} \textbf{\bibinfo{volume}{82}},
  \bibinfo{pages}{012339} (\bibinfo{year}{2010}).

\bibitem[{\citenamefont{Goerz et~al.}(2014{\natexlab{a}})\citenamefont{Goerz,
  Halperin, Aytac, Koch, and Whaley}}]{goerzRobustnessHighfidelityRydberg2014}
\bibinfo{author}{\bibfnamefont{M.~H.} \bibnamefont{Goerz}},
  \bibinfo{author}{\bibfnamefont{E.~J.} \bibnamefont{Halperin}},
  \bibinfo{author}{\bibfnamefont{J.~M.} \bibnamefont{Aytac}},
  \bibinfo{author}{\bibfnamefont{C.~P.} \bibnamefont{Koch}}, \bibnamefont{and}
  \bibinfo{author}{\bibfnamefont{K.~B.} \bibnamefont{Whaley}},
  \bibinfo{journal}{Physical Review A} \textbf{\bibinfo{volume}{90}},
  \bibinfo{pages}{032329} (\bibinfo{year}{2014}{\natexlab{a}}).

\bibitem[{\citenamefont{Treutlein et~al.}(2006)\citenamefont{Treutlein,
  H{\"a}nsch, Reichel, Negretti, Cirone, and
  Calarco}}]{treutleinMicrowavePotentialsOptimal2006}
\bibinfo{author}{\bibfnamefont{P.}~\bibnamefont{Treutlein}},
  \bibinfo{author}{\bibfnamefont{T.~W.} \bibnamefont{H{\"a}nsch}},
  \bibinfo{author}{\bibfnamefont{J.}~\bibnamefont{Reichel}},
  \bibinfo{author}{\bibfnamefont{A.}~\bibnamefont{Negretti}},
  \bibinfo{author}{\bibfnamefont{M.~A.} \bibnamefont{Cirone}},
  \bibnamefont{and} \bibinfo{author}{\bibfnamefont{T.}~\bibnamefont{Calarco}},
  \bibinfo{journal}{Physical Review A} \textbf{\bibinfo{volume}{74}},
  \bibinfo{pages}{022312} (\bibinfo{year}{2006}).

\bibitem[{\citenamefont{Gorshkov et~al.}(2008)\citenamefont{Gorshkov, Calarco,
  Lukin, and S{\o}rensen}}]{gorshkovPhotonStorageEnsuremath2008}
\bibinfo{author}{\bibfnamefont{A.~V.} \bibnamefont{Gorshkov}},
  \bibinfo{author}{\bibfnamefont{T.}~\bibnamefont{Calarco}},
  \bibinfo{author}{\bibfnamefont{M.~D.} \bibnamefont{Lukin}}, \bibnamefont{and}
  \bibinfo{author}{\bibfnamefont{A.~S.} \bibnamefont{S{\o}rensen}},
  \bibinfo{journal}{Physical Review A} \textbf{\bibinfo{volume}{77}},
  \bibinfo{pages}{043806} (\bibinfo{year}{2008}).

\bibitem[{\citenamefont{Goerz et~al.}(2014{\natexlab{b}})\citenamefont{Goerz,
  Reich, and Koch}}]{goerzOptimalControlTheory2014}
\bibinfo{author}{\bibfnamefont{M.~H.} \bibnamefont{Goerz}},
  \bibinfo{author}{\bibfnamefont{D.~M.} \bibnamefont{Reich}}, \bibnamefont{and}
  \bibinfo{author}{\bibfnamefont{C.~P.} \bibnamefont{Koch}},
  \bibinfo{journal}{New Journal of Physics} \textbf{\bibinfo{volume}{16}},
  \bibinfo{pages}{055012} (\bibinfo{year}{2014}{\natexlab{b}}), ISSN
  \bibinfo{issn}{1367-2630}.

\bibitem[{\citenamefont{Werninghaus et~al.}(2021)\citenamefont{Werninghaus,
  Egger, Roy, Machnes, Wilhelm, and
  Filipp}}]{werninghausLeakageReductionFast2021}
\bibinfo{author}{\bibfnamefont{M.}~\bibnamefont{Werninghaus}},
  \bibinfo{author}{\bibfnamefont{D.~J.} \bibnamefont{Egger}},
  \bibinfo{author}{\bibfnamefont{F.}~\bibnamefont{Roy}},
  \bibinfo{author}{\bibfnamefont{S.}~\bibnamefont{Machnes}},
  \bibinfo{author}{\bibfnamefont{F.~K.} \bibnamefont{Wilhelm}},
  \bibnamefont{and} \bibinfo{author}{\bibfnamefont{S.}~\bibnamefont{Filipp}},
  \bibinfo{journal}{npj Quantum Information} \textbf{\bibinfo{volume}{7}},
  \bibinfo{pages}{1} (\bibinfo{year}{2021}), ISSN \bibinfo{issn}{2056-6387}.

\bibitem[{\citenamefont{Abdelhafez et~al.}(2020)\citenamefont{Abdelhafez,
  Baker, Gyenis, Mundada, Houck, Schuster, and
  Koch}}]{abdelhafezUniversalGatesProtected2020}
\bibinfo{author}{\bibfnamefont{M.}~\bibnamefont{Abdelhafez}},
  \bibinfo{author}{\bibfnamefont{B.}~\bibnamefont{Baker}},
  \bibinfo{author}{\bibfnamefont{A.}~\bibnamefont{Gyenis}},
  \bibinfo{author}{\bibfnamefont{P.}~\bibnamefont{Mundada}},
  \bibinfo{author}{\bibfnamefont{A.~A.} \bibnamefont{Houck}},
  \bibinfo{author}{\bibfnamefont{D.}~\bibnamefont{Schuster}}, \bibnamefont{and}
  \bibinfo{author}{\bibfnamefont{J.}~\bibnamefont{Koch}},
  \bibinfo{journal}{Physical Review A} \textbf{\bibinfo{volume}{101}},
  \bibinfo{pages}{022321} (\bibinfo{year}{2020}).

\bibitem[{\citenamefont{Scheuer et~al.}(2014)\citenamefont{Scheuer, Kong, Said,
  Chen, Kurz, Marseglia, Du, Hemmer, Montangero, Calarco
  et~al.}}]{scheuerPreciseQubitControl2014}
\bibinfo{author}{\bibfnamefont{J.}~\bibnamefont{Scheuer}},
  \bibinfo{author}{\bibfnamefont{X.}~\bibnamefont{Kong}},
  \bibinfo{author}{\bibfnamefont{R.~S.} \bibnamefont{Said}},
  \bibinfo{author}{\bibfnamefont{J.}~\bibnamefont{Chen}},
  \bibinfo{author}{\bibfnamefont{A.}~\bibnamefont{Kurz}},
  \bibinfo{author}{\bibfnamefont{L.}~\bibnamefont{Marseglia}},
  \bibinfo{author}{\bibfnamefont{J.}~\bibnamefont{Du}},
  \bibinfo{author}{\bibfnamefont{P.~R.} \bibnamefont{Hemmer}},
  \bibinfo{author}{\bibfnamefont{S.}~\bibnamefont{Montangero}},
  \bibinfo{author}{\bibfnamefont{T.}~\bibnamefont{Calarco}},
  \bibnamefont{et~al.}, \bibinfo{journal}{New Journal of Physics}
  \textbf{\bibinfo{volume}{16}}, \bibinfo{pages}{093022}
  (\bibinfo{year}{2014}), ISSN \bibinfo{issn}{1367-2630}.

\bibitem[{\citenamefont{Waldherr et~al.}(2014)\citenamefont{Waldherr, Wang,
  Zaiser, Jamali, {Schulte-Herbr{\"u}ggen}, Abe, Ohshima, Isoya, Du, Neumann
  et~al.}}]{waldherrQuantumErrorCorrection2014}
\bibinfo{author}{\bibfnamefont{G.}~\bibnamefont{Waldherr}},
  \bibinfo{author}{\bibfnamefont{Y.}~\bibnamefont{Wang}},
  \bibinfo{author}{\bibfnamefont{S.}~\bibnamefont{Zaiser}},
  \bibinfo{author}{\bibfnamefont{M.}~\bibnamefont{Jamali}},
  \bibinfo{author}{\bibfnamefont{T.}~\bibnamefont{{Schulte-Herbr{\"u}ggen}}},
  \bibinfo{author}{\bibfnamefont{H.}~\bibnamefont{Abe}},
  \bibinfo{author}{\bibfnamefont{T.}~\bibnamefont{Ohshima}},
  \bibinfo{author}{\bibfnamefont{J.}~\bibnamefont{Isoya}},
  \bibinfo{author}{\bibfnamefont{J.~F.} \bibnamefont{Du}},
  \bibinfo{author}{\bibfnamefont{P.}~\bibnamefont{Neumann}},
  \bibnamefont{et~al.}, \bibinfo{journal}{Nature}
  \textbf{\bibinfo{volume}{506}}, \bibinfo{pages}{204} (\bibinfo{year}{2014}),
  ISSN \bibinfo{issn}{1476-4687}.

\bibitem[{\citenamefont{Li and
  Khaneja}(2006)}]{liControlInhomogeneousQuantum2006}
\bibinfo{author}{\bibfnamefont{J.-S.} \bibnamefont{Li}} \bibnamefont{and}
  \bibinfo{author}{\bibfnamefont{N.}~\bibnamefont{Khaneja}},
  \bibinfo{journal}{Physical Review A} \textbf{\bibinfo{volume}{73}},
  \bibinfo{pages}{030302} (\bibinfo{year}{2006}).

\bibitem[{\citenamefont{Li and Khaneja}(2009)}]{liEnsembleControlBloch2009}
\bibinfo{author}{\bibfnamefont{J.-S.} \bibnamefont{Li}} \bibnamefont{and}
  \bibinfo{author}{\bibfnamefont{N.}~\bibnamefont{Khaneja}},
  \bibinfo{journal}{IEEE Transactions on Automatic Control}
  \textbf{\bibinfo{volume}{54}}, \bibinfo{pages}{528} (\bibinfo{year}{2009}),
  ISSN \bibinfo{issn}{1558-2523}.

\bibitem[{\citenamefont{Rabitz and
  Turinici}(2007)}]{rabitzControllingQuantumDynamics2007}
\bibinfo{author}{\bibfnamefont{H.}~\bibnamefont{Rabitz}} \bibnamefont{and}
  \bibinfo{author}{\bibfnamefont{G.}~\bibnamefont{Turinici}},
  \bibinfo{journal}{Physical Review A} \textbf{\bibinfo{volume}{75}},
  \bibinfo{pages}{043409} (\bibinfo{year}{2007}).

\bibitem[{\citenamefont{Turinici and
  Rabitz}(2004)}]{turiniciOptimallyControllingInternal2004}
\bibinfo{author}{\bibfnamefont{G.}~\bibnamefont{Turinici}} \bibnamefont{and}
  \bibinfo{author}{\bibfnamefont{H.}~\bibnamefont{Rabitz}},
  \bibinfo{journal}{Physical Review A} \textbf{\bibinfo{volume}{70}},
  \bibinfo{pages}{063412} (\bibinfo{year}{2004}), ISSN
  \bibinfo{issn}{1050-2947, 1094-1622}.

\bibitem[{\citenamefont{Cummins and
  Jones}(2000)}]{cumminsUseCompositeRotations2000}
\bibinfo{author}{\bibfnamefont{H.~K.} \bibnamefont{Cummins}} \bibnamefont{and}
  \bibinfo{author}{\bibfnamefont{J.~A.} \bibnamefont{Jones}},
  \bibinfo{journal}{New Journal of Physics} \textbf{\bibinfo{volume}{2}},
  \bibinfo{pages}{6} (\bibinfo{year}{2000}), ISSN \bibinfo{issn}{1367-2630}.

\bibitem[{\citenamefont{Brown et~al.}(2004)\citenamefont{Brown, Harrow, and
  Chuang}}]{brownArbitrarilyAccurateComposite2004}
\bibinfo{author}{\bibfnamefont{K.~R.} \bibnamefont{Brown}},
  \bibinfo{author}{\bibfnamefont{A.~W.} \bibnamefont{Harrow}},
  \bibnamefont{and} \bibinfo{author}{\bibfnamefont{I.~L.}
  \bibnamefont{Chuang}}, \bibinfo{journal}{Physical Review A}
  \textbf{\bibinfo{volume}{70}}, \bibinfo{pages}{052318}
  (\bibinfo{year}{2004}), ISSN \bibinfo{issn}{1050-2947, 1094-1622}.

\bibitem[{\citenamefont{Owrutsky and
  Khaneja}(2012)}]{owrutskyControlInhomogeneousEnsembles2012}
\bibinfo{author}{\bibfnamefont{P.}~\bibnamefont{Owrutsky}} \bibnamefont{and}
  \bibinfo{author}{\bibfnamefont{N.}~\bibnamefont{Khaneja}},
  \bibinfo{journal}{Physical Review A} \textbf{\bibinfo{volume}{86}},
  \bibinfo{pages}{022315} (\bibinfo{year}{2012}).

\bibitem[{\citenamefont{Ansel et~al.}(2021)\citenamefont{Ansel, Glaser, and
  Sugny}}]{anselSelectiveRobustTimeoptimal2021}
\bibinfo{author}{\bibfnamefont{Q.}~\bibnamefont{Ansel}},
  \bibinfo{author}{\bibfnamefont{S.~J.} \bibnamefont{Glaser}},
  \bibnamefont{and} \bibinfo{author}{\bibfnamefont{D.}~\bibnamefont{Sugny}},
  \bibinfo{journal}{Journal of Physics A: Mathematical and Theoretical}
  \textbf{\bibinfo{volume}{54}}, \bibinfo{pages}{085204}
  (\bibinfo{year}{2021}), ISSN \bibinfo{issn}{1751-8121}.

\bibitem[{\citenamefont{Augier et~al.}(2018)\citenamefont{Augier, Boscain, and
  Sigalotti}}]{augierAdiabaticEnsembleControl2018}
\bibinfo{author}{\bibfnamefont{N.}~\bibnamefont{Augier}},
  \bibinfo{author}{\bibfnamefont{U.}~\bibnamefont{Boscain}}, \bibnamefont{and}
  \bibinfo{author}{\bibfnamefont{M.}~\bibnamefont{Sigalotti}},
  \bibinfo{journal}{SIAM Journal on Control and Optimization}
  \textbf{\bibinfo{volume}{56}}, \bibinfo{pages}{4045} (\bibinfo{year}{2018}),
  ISSN \bibinfo{issn}{0363-0129}.

\bibitem[{\citenamefont{Tycko}(1983)}]{tyckoBroadbandPopulationInversion1983}
\bibinfo{author}{\bibfnamefont{R.}~\bibnamefont{Tycko}},
  \bibinfo{journal}{Physical Review Letters} \textbf{\bibinfo{volume}{51}},
  \bibinfo{pages}{775} (\bibinfo{year}{1983}), ISSN \bibinfo{issn}{0031-9007}.

\bibitem[{\citenamefont{Levitt}(1986)}]{levittCompositePulses1986}
\bibinfo{author}{\bibfnamefont{M.~H.} \bibnamefont{Levitt}},
  \bibinfo{journal}{Progress in Nuclear Magnetic Resonance Spectroscopy}
  \textbf{\bibinfo{volume}{18}}, \bibinfo{pages}{61} (\bibinfo{year}{1986}),
  ISSN \bibinfo{issn}{0079-6565}.

\bibitem[{\citenamefont{Mischuck et~al.}(2012)\citenamefont{Mischuck, Merkel,
  and Deutsch}}]{mischuckControlInhomogeneousAtomic2012}
\bibinfo{author}{\bibfnamefont{B.~E.} \bibnamefont{Mischuck}},
  \bibinfo{author}{\bibfnamefont{S.~T.} \bibnamefont{Merkel}},
  \bibnamefont{and} \bibinfo{author}{\bibfnamefont{I.~H.}
  \bibnamefont{Deutsch}}, \bibinfo{journal}{Physical Review A}
  \textbf{\bibinfo{volume}{85}}, \bibinfo{pages}{022302}
  (\bibinfo{year}{2012}).

\bibitem[{\citenamefont{Khani et~al.}(2012)\citenamefont{Khani, Merkel, Motzoi,
  Gambetta, and Wilhelm}}]{khaniHighfidelityQuantumGates2012}
\bibinfo{author}{\bibfnamefont{B.}~\bibnamefont{Khani}},
  \bibinfo{author}{\bibfnamefont{S.~T.} \bibnamefont{Merkel}},
  \bibinfo{author}{\bibfnamefont{F.}~\bibnamefont{Motzoi}},
  \bibinfo{author}{\bibfnamefont{J.~M.} \bibnamefont{Gambetta}},
  \bibnamefont{and} \bibinfo{author}{\bibfnamefont{F.~K.}
  \bibnamefont{Wilhelm}}, \bibinfo{journal}{Physical Review A}
  \textbf{\bibinfo{volume}{85}}, \bibinfo{pages}{022306}
  (\bibinfo{year}{2012}).

\bibitem[{\citenamefont{Khaneja et~al.}(2005)\citenamefont{Khaneja, Reiss,
  Kehlet, {Schulte-Herbr{\"u}ggen}, and
  Glaser}}]{khanejaOptimalControlCoupled2005a}
\bibinfo{author}{\bibfnamefont{N.}~\bibnamefont{Khaneja}},
  \bibinfo{author}{\bibfnamefont{T.}~\bibnamefont{Reiss}},
  \bibinfo{author}{\bibfnamefont{C.}~\bibnamefont{Kehlet}},
  \bibinfo{author}{\bibfnamefont{T.}~\bibnamefont{{Schulte-Herbr{\"u}ggen}}},
  \bibnamefont{and} \bibinfo{author}{\bibfnamefont{S.~J.}
  \bibnamefont{Glaser}}, \bibinfo{journal}{Journal of Magnetic Resonance}
  \textbf{\bibinfo{volume}{172}}, \bibinfo{pages}{296} (\bibinfo{year}{2005}),
  ISSN \bibinfo{issn}{1090-7807}.

\bibitem[{\citenamefont{Ruths and
  Li}(2011)}]{ruthsMultidimensionalPseudospectralMethod2011}
\bibinfo{author}{\bibfnamefont{J.}~\bibnamefont{Ruths}} \bibnamefont{and}
  \bibinfo{author}{\bibfnamefont{J.-S.} \bibnamefont{Li}},
  \bibinfo{journal}{The Journal of Chemical Physics}
  \textbf{\bibinfo{volume}{134}}, \bibinfo{pages}{044128}
  (\bibinfo{year}{2011}), ISSN \bibinfo{issn}{0021-9606}.

\bibitem[{\citenamefont{Chen et~al.}(2014)\citenamefont{Chen, Dong, Long,
  Petersen, and Rabitz}}]{chenSamplingbasedLearningControl2014}
\bibinfo{author}{\bibfnamefont{C.}~\bibnamefont{Chen}},
  \bibinfo{author}{\bibfnamefont{D.}~\bibnamefont{Dong}},
  \bibinfo{author}{\bibfnamefont{R.}~\bibnamefont{Long}},
  \bibinfo{author}{\bibfnamefont{I.~R.} \bibnamefont{Petersen}},
  \bibnamefont{and} \bibinfo{author}{\bibfnamefont{H.~A.}
  \bibnamefont{Rabitz}}, \bibinfo{journal}{Physical Review A}
  \textbf{\bibinfo{volume}{89}}, \bibinfo{pages}{023402}
  (\bibinfo{year}{2014}), ISSN \bibinfo{issn}{1050-2947, 1094-1622}.

\bibitem[{\citenamefont{Li et~al.}(2011)\citenamefont{Li, Ruths, Yu, Arthanari,
  and Wagner}}]{liOptimalPulseDesign2011}
\bibinfo{author}{\bibfnamefont{J.-S.} \bibnamefont{Li}},
  \bibinfo{author}{\bibfnamefont{J.}~\bibnamefont{Ruths}},
  \bibinfo{author}{\bibfnamefont{T.-Y.} \bibnamefont{Yu}},
  \bibinfo{author}{\bibfnamefont{H.}~\bibnamefont{Arthanari}},
  \bibnamefont{and} \bibinfo{author}{\bibfnamefont{G.}~\bibnamefont{Wagner}},
  \bibinfo{journal}{Proceedings of the National Academy of Sciences}
  \textbf{\bibinfo{volume}{108}}, \bibinfo{pages}{1879} (\bibinfo{year}{2011}),
  ISSN \bibinfo{issn}{0027-8424, 1091-6490}.

\bibitem[{\citenamefont{Ruths and
  Li}(2012)}]{ruthsOptimalControlInhomogeneous2012}
\bibinfo{author}{\bibfnamefont{J.}~\bibnamefont{Ruths}} \bibnamefont{and}
  \bibinfo{author}{\bibfnamefont{J.-S.} \bibnamefont{Li}},
  \bibinfo{journal}{IEEE Transactions on Automatic Control}
  \textbf{\bibinfo{volume}{57}}, \bibinfo{pages}{2021} (\bibinfo{year}{2012}),
  ISSN \bibinfo{issn}{1558-2523}.

\bibitem[{\citenamefont{Turinici}(2019)}]{turiniciStochasticLearningControl2019}
\bibinfo{author}{\bibfnamefont{G.}~\bibnamefont{Turinici}},
  \bibinfo{journal}{Physical Review A} \textbf{\bibinfo{volume}{100}},
  \bibinfo{pages}{053403} (\bibinfo{year}{2019}), ISSN
  \bibinfo{issn}{2469-9926, 2469-9934}.

\bibitem[{\citenamefont{Kuang and
  Guan}(2020)}]{kuangRobustnessContinuousNonsmooth2020}
\bibinfo{author}{\bibfnamefont{S.}~\bibnamefont{Kuang}} \bibnamefont{and}
  \bibinfo{author}{\bibfnamefont{X.}~\bibnamefont{Guan}}, \bibinfo{journal}{IET
  Control Theory \& Applications} \textbf{\bibinfo{volume}{14}},
  \bibinfo{pages}{2449} (\bibinfo{year}{2020}), ISSN \bibinfo{issn}{1751-8652}.

\bibitem[{\citenamefont{Wu et~al.}(2019)\citenamefont{Wu, Ding, Dong, and
  Wang}}]{wuLearningRobustHighprecision2019}
\bibinfo{author}{\bibfnamefont{R.-B.} \bibnamefont{Wu}},
  \bibinfo{author}{\bibfnamefont{H.}~\bibnamefont{Ding}},
  \bibinfo{author}{\bibfnamefont{D.}~\bibnamefont{Dong}}, \bibnamefont{and}
  \bibinfo{author}{\bibfnamefont{X.}~\bibnamefont{Wang}},
  \bibinfo{journal}{Physical Review A} \textbf{\bibinfo{volume}{99}},
  \bibinfo{pages}{042327} (\bibinfo{year}{2019}), ISSN
  \bibinfo{issn}{2469-9926, 2469-9934}.

\bibitem[{\citenamefont{Sun et~al.}(2015)\citenamefont{Sun, Ma, Wu, Chen, and
  Dong}}]{sunEnsembleControlOpen2015}
\bibinfo{author}{\bibfnamefont{Y.}~\bibnamefont{Sun}},
  \bibinfo{author}{\bibfnamefont{H.}~\bibnamefont{Ma}},
  \bibinfo{author}{\bibfnamefont{C.}~\bibnamefont{Wu}},
  \bibinfo{author}{\bibfnamefont{C.}~\bibnamefont{Chen}}, \bibnamefont{and}
  \bibinfo{author}{\bibfnamefont{D.}~\bibnamefont{Dong}}, in
  \emph{\bibinfo{booktitle}{2015 10th {{Asian Control Conference}} ({{ASCC}})}}
  (\bibinfo{year}{2015}), pp. \bibinfo{pages}{1--6}.

\bibitem[{\citenamefont{Wang and
  Li}(2018)}]{wangFreeendpointOptimalControl2018}
\bibinfo{author}{\bibfnamefont{S.}~\bibnamefont{Wang}} \bibnamefont{and}
  \bibinfo{author}{\bibfnamefont{J.-S.} \bibnamefont{Li}},
  \bibinfo{journal}{Automatica} \textbf{\bibinfo{volume}{95}},
  \bibinfo{pages}{306} (\bibinfo{year}{2018}), ISSN \bibinfo{issn}{0005-1098}.

\bibitem[{\citenamefont{Kuang et~al.}(2018)\citenamefont{Kuang, Qi, and
  Cong}}]{kuangApproximateTimeoptimalControl2018}
\bibinfo{author}{\bibfnamefont{S.}~\bibnamefont{Kuang}},
  \bibinfo{author}{\bibfnamefont{P.}~\bibnamefont{Qi}}, \bibnamefont{and}
  \bibinfo{author}{\bibfnamefont{S.}~\bibnamefont{Cong}},
  \bibinfo{journal}{Physics Letters A} \textbf{\bibinfo{volume}{382}},
  \bibinfo{pages}{1858} (\bibinfo{year}{2018}), ISSN \bibinfo{issn}{0375-9601}.

\bibitem[{\citenamefont{Van~Damme et~al.}(2017)\citenamefont{Van~Damme, Ansel,
  Glaser, and Sugny}}]{vandammeRobustOptimalControl2017}
\bibinfo{author}{\bibfnamefont{L.}~\bibnamefont{Van~Damme}},
  \bibinfo{author}{\bibfnamefont{Q.}~\bibnamefont{Ansel}},
  \bibinfo{author}{\bibfnamefont{S.~J.} \bibnamefont{Glaser}},
  \bibnamefont{and} \bibinfo{author}{\bibfnamefont{D.}~\bibnamefont{Sugny}},
  \bibinfo{journal}{Physical Review A} \textbf{\bibinfo{volume}{95}},
  \bibinfo{pages}{063403} (\bibinfo{year}{2017}), ISSN
  \bibinfo{issn}{2469-9926, 2469-9934}.

\bibitem[{\citenamefont{Arjmandzadeh and
  Yarahmadi}(2017)}]{arjmandzadehQuantumGeneticLearning2017}
\bibinfo{author}{\bibfnamefont{A.}~\bibnamefont{Arjmandzadeh}}
  \bibnamefont{and}
  \bibinfo{author}{\bibfnamefont{M.}~\bibnamefont{Yarahmadi}},
  \bibinfo{journal}{Entropy} \textbf{\bibinfo{volume}{19}},
  \bibinfo{pages}{376} (\bibinfo{year}{2017}).

\bibitem[{\citenamefont{Trivedi et~al.}(2018)\citenamefont{Trivedi, Fischer,
  Xu, Fan, and Vuckovic}}]{trivediFewphotonScatteringEmission2018}
\bibinfo{author}{\bibfnamefont{R.}~\bibnamefont{Trivedi}},
  \bibinfo{author}{\bibfnamefont{K.}~\bibnamefont{Fischer}},
  \bibinfo{author}{\bibfnamefont{S.}~\bibnamefont{Xu}},
  \bibinfo{author}{\bibfnamefont{S.}~\bibnamefont{Fan}}, \bibnamefont{and}
  \bibinfo{author}{\bibfnamefont{J.}~\bibnamefont{Vuckovic}},
  \bibinfo{journal}{Physical Review B} \textbf{\bibinfo{volume}{98}},
  \bibinfo{pages}{144112} (\bibinfo{year}{2018}).

\bibitem[{\citenamefont{Trivedi
  et~al.}(2020{\natexlab{a}})\citenamefont{Trivedi, White, Fan, and Vu{\v
  c}kovi{\'c}}}]{trivediAnalyticGeometricProperties2020b}
\bibinfo{author}{\bibfnamefont{R.}~\bibnamefont{Trivedi}},
  \bibinfo{author}{\bibfnamefont{A.}~\bibnamefont{White}},
  \bibinfo{author}{\bibfnamefont{S.}~\bibnamefont{Fan}}, \bibnamefont{and}
  \bibinfo{author}{\bibfnamefont{J.}~\bibnamefont{Vu{\v c}kovi{\'c}}},
  \bibinfo{journal}{Physical Review A} \textbf{\bibinfo{volume}{102}},
  \bibinfo{pages}{033707} (\bibinfo{year}{2020}{\natexlab{a}}).

\bibitem[{\citenamefont{Rephaeli and
  Fan}(2012)}]{rephaeliFewPhotonSingleAtomCavity2012}
\bibinfo{author}{\bibfnamefont{E.}~\bibnamefont{Rephaeli}} \bibnamefont{and}
  \bibinfo{author}{\bibfnamefont{S.}~\bibnamefont{Fan}}, \bibinfo{journal}{IEEE
  Journal of Selected Topics in Quantum Electronics}
  \textbf{\bibinfo{volume}{18}}, \bibinfo{pages}{1754} (\bibinfo{year}{2012}),
  ISSN \bibinfo{issn}{1558-4542}.

\bibitem[{\citenamefont{Fan et~al.}(2010)\citenamefont{Fan, Kocaba{\c s}, and
  Shen}}]{fanInputoutputFormalismFewphoton2010}
\bibinfo{author}{\bibfnamefont{S.}~\bibnamefont{Fan}},
  \bibinfo{author}{\bibfnamefont{{\c S}.~E.} \bibnamefont{Kocaba{\c s}}},
  \bibnamefont{and} \bibinfo{author}{\bibfnamefont{J.-T.} \bibnamefont{Shen}},
  \bibinfo{journal}{Physical Review A} \textbf{\bibinfo{volume}{82}},
  \bibinfo{pages}{063821} (\bibinfo{year}{2010}).

\bibitem[{\citenamefont{Schmidt}(2006)}]{schmidtNumericalMethodsOptimal2006}
\bibinfo{author}{\bibfnamefont{W.~H.} \bibnamefont{Schmidt}}, in
  \emph{\bibinfo{booktitle}{Large-{{Scale Scientific Computing}}}}, edited by
  \bibinfo{editor}{\bibfnamefont{I.}~\bibnamefont{Lirkov}},
  \bibinfo{editor}{\bibfnamefont{S.}~\bibnamefont{Margenov}}, \bibnamefont{and}
  \bibinfo{editor}{\bibfnamefont{J.}~\bibnamefont{Wa{\'s}niewski}}
  (\bibinfo{publisher}{{Springer}}, \bibinfo{address}{{Berlin, Heidelberg}},
  \bibinfo{year}{2006}), Lecture {{Notes}} in {{Computer Science}}, pp.
  \bibinfo{pages}{255--262}, ISBN \bibinfo{isbn}{978-3-540-31995-5}.

\bibitem[{\citenamefont{Swillam et~al.}(2007)\citenamefont{Swillam, Bakr, and
  Li}}]{swillamAccurateSensitivityAnalysis2007}
\bibinfo{author}{\bibfnamefont{M.~A.} \bibnamefont{Swillam}},
  \bibinfo{author}{\bibfnamefont{M.~H.} \bibnamefont{Bakr}}, \bibnamefont{and}
  \bibinfo{author}{\bibfnamefont{X.}~\bibnamefont{Li}},
  \bibinfo{journal}{Applied Optics} \textbf{\bibinfo{volume}{46}},
  \bibinfo{pages}{1492} (\bibinfo{year}{2007}), ISSN \bibinfo{issn}{2155-3165}.

\bibitem[{\citenamefont{Scott}(1992)}]{scottMultivariateDensityEstimation1992}
\bibinfo{author}{\bibfnamefont{D.~W.} \bibnamefont{Scott}},
  \emph{\bibinfo{title}{Multivariate {{Density Estimation}}}}
  (\bibinfo{publisher}{{John Wiley \& Sons, Ltd}}, \bibinfo{year}{1992}),
  \bibinfo{edition}{1st} ed.

\bibitem[{\citenamefont{Boyd and
  Vandenberghe}(2004)}]{boydConvexOptimization2004}
\bibinfo{author}{\bibfnamefont{S.}~\bibnamefont{Boyd}} \bibnamefont{and}
  \bibinfo{author}{\bibfnamefont{L.}~\bibnamefont{Vandenberghe}},
  \emph{\bibinfo{title}{Convex {{Optimization}}}}
  (\bibinfo{publisher}{{Cambridge University Press}},
  \bibinfo{address}{{Cambridge}}, \bibinfo{year}{2004}), ISBN
  \bibinfo{isbn}{978-0-521-83378-3}.

\bibitem[{\citenamefont{Trivedi
  et~al.}(2020{\natexlab{b}})\citenamefont{Trivedi, Angeris, Su, Boyd, Fan, and
  Vu{\v c}kovi{\'c}}}]{trivediBoundsScatteringAbsorptionless2020}
\bibinfo{author}{\bibfnamefont{R.}~\bibnamefont{Trivedi}},
  \bibinfo{author}{\bibfnamefont{G.}~\bibnamefont{Angeris}},
  \bibinfo{author}{\bibfnamefont{L.}~\bibnamefont{Su}},
  \bibinfo{author}{\bibfnamefont{S.}~\bibnamefont{Boyd}},
  \bibinfo{author}{\bibfnamefont{S.}~\bibnamefont{Fan}}, \bibnamefont{and}
  \bibinfo{author}{\bibfnamefont{J.}~\bibnamefont{Vu{\v c}kovi{\'c}}},
  \bibinfo{journal}{Physical Review Applied} \textbf{\bibinfo{volume}{14}},
  \bibinfo{pages}{014025} (\bibinfo{year}{2020}{\natexlab{b}}).

\bibitem[{\citenamefont{Debnath et~al.}(2019)\citenamefont{Debnath, Zhang, and
  M{\o}lmer}}]{debnathCollectiveDynamicsInhomogeneously2019a}
\bibinfo{author}{\bibfnamefont{K.}~\bibnamefont{Debnath}},
  \bibinfo{author}{\bibfnamefont{Y.}~\bibnamefont{Zhang}}, \bibnamefont{and}
  \bibinfo{author}{\bibfnamefont{K.}~\bibnamefont{M{\o}lmer}},
  \bibinfo{journal}{Physical Review A} \textbf{\bibinfo{volume}{100}},
  \bibinfo{pages}{053821} (\bibinfo{year}{2019}).

\bibitem[{\citenamefont{Lukin et~al.}(2020{\natexlab{b}})\citenamefont{Lukin,
  White, Trivedi, Guidry, Morioka, Babin, Soykal, {Ul-Hassan}, Son, Ohshima
  et~al.}}]{lukinSpectrallyReconfigurableQuantum2020}
\bibinfo{author}{\bibfnamefont{D.~M.} \bibnamefont{Lukin}},
  \bibinfo{author}{\bibfnamefont{A.~D.} \bibnamefont{White}},
  \bibinfo{author}{\bibfnamefont{R.}~\bibnamefont{Trivedi}},
  \bibinfo{author}{\bibfnamefont{M.~A.} \bibnamefont{Guidry}},
  \bibinfo{author}{\bibfnamefont{N.}~\bibnamefont{Morioka}},
  \bibinfo{author}{\bibfnamefont{C.}~\bibnamefont{Babin}},
  \bibinfo{author}{\bibfnamefont{{\"O}.~O.} \bibnamefont{Soykal}},
  \bibinfo{author}{\bibfnamefont{J.}~\bibnamefont{{Ul-Hassan}}},
  \bibinfo{author}{\bibfnamefont{N.~T.} \bibnamefont{Son}},
  \bibinfo{author}{\bibfnamefont{T.}~\bibnamefont{Ohshima}},
  \bibnamefont{et~al.}, \bibinfo{journal}{npj Quantum Information}
  \textbf{\bibinfo{volume}{6}}, \bibinfo{pages}{1}
  (\bibinfo{year}{2020}{\natexlab{b}}), ISSN \bibinfo{issn}{2056-6387}.

\bibitem[{\citenamefont{Kingma and Ba}(2017)}]{kingmaAdamMethodStochastic2017}
\bibinfo{author}{\bibfnamefont{D.~P.} \bibnamefont{Kingma}} \bibnamefont{and}
  \bibinfo{author}{\bibfnamefont{J.}~\bibnamefont{Ba}},
  \bibinfo{journal}{arXiv:1412.6980 [cs]}  (\bibinfo{year}{2017}),
  \bibinfo{note}{comment: Published as a conference paper at the 3rd
  International Conference for Learning Representations, San Diego, 2015},
  \eprint{1412.6980}.

\bibitem[{\citenamefont{Brecht et~al.}(2015)\citenamefont{Brecht, Reddy,
  Silberhorn, and Raymer}}]{brechtPhotonTemporalModes2015}
\bibinfo{author}{\bibfnamefont{B.}~\bibnamefont{Brecht}},
  \bibinfo{author}{\bibfnamefont{D.~V.} \bibnamefont{Reddy}},
  \bibinfo{author}{\bibfnamefont{C.}~\bibnamefont{Silberhorn}},
  \bibnamefont{and} \bibinfo{author}{\bibfnamefont{M.~G.}
  \bibnamefont{Raymer}}, \bibinfo{journal}{Physical Review X}
  \textbf{\bibinfo{volume}{5}}, \bibinfo{pages}{041017} (\bibinfo{year}{2015}),
  ISSN \bibinfo{issn}{2160-3308}.

\bibitem[{\citenamefont{Courant and
  Hilbert}(1989)}]{courantSeriesExpansionsArbitrary1989}
\bibinfo{author}{\bibfnamefont{R.}~\bibnamefont{Courant}} \bibnamefont{and}
  \bibinfo{author}{\bibfnamefont{D.}~\bibnamefont{Hilbert}}, in
  \emph{\bibinfo{booktitle}{Methods of {{Mathematical Physics}}}}
  (\bibinfo{publisher}{{John Wiley \& Sons, Ltd}}, \bibinfo{year}{1989}),
  chap.~\bibinfo{chapter}{2}, pp. \bibinfo{pages}{48--111}, ISBN
  \bibinfo{isbn}{978-3-527-61721-0}.

\end{thebibliography}

\end{document}